\documentclass[10pt,aps,onecolumn,superscriptaddress]{revtex4-2}

\usepackage[T1]{fontenc}
\usepackage[utf8]{inputenc}
\usepackage[english]{babel}

\usepackage{amsfonts,amsbsy,amssymb,amsmath}
\usepackage{graphicx,float}
\usepackage{hyperref}
\hypersetup{
    colorlinks=true,
    linkcolor=blue,
    filecolor=magenta,      
    urlcolor=cyan,
    citecolor=blue,
}

\graphicspath{{notebooks/figures/}}

\begin{document}

\title{Transport and Roaming on the Double van der Waals Potential Energy Surface}

\author{Francisco Gonz\'alez Montoya}
\email{francisco.glz.mty@gmail.com}
\affiliation{School of Mathematics, University of Bristol, \\ Fry Building, Woodland Road, Bristol, BS8 1UG, United Kingdom.}
\author{V\'ictor J. Garc\'ia-Garrido}
\email{vjose.garcia@uah.es}
\affiliation{Departamento de F\'isica y Matem\'aticas, Universidad de Alcal\'a,\\ Madrid, 28871, Spain.}
\author{Broncio Aguilar-Sanjuan}
\email{ba13026@bristol.ac.uk}
\affiliation{School of Mathematics, University of Bristol, \\ Fry Building, Woodland Road, Bristol, BS8 1UG, United Kingdom.}
\author{Stephen Wiggins}
\email{s.wiggins@bristol.ac.uk}
\affiliation{School of Mathematics, University of Bristol, \\ Fry Building, Woodland Road, Bristol, BS8 1UG, United Kingdom.}

\begin{abstract}

This paper explores the phase space structures characterising transport for a double-well van der Waals potential surface. Trajectories are classified as inter-well, intra-well, and escaping-from-a-well to define different dynamical fates. In particular, roaming trajectories, which are a new paradigm in chemical reaction dynamics, are observed. We apply the method of Lagrangian descriptors to uncover the boundaries between the fates. We find that the boundaries between different fates are the stable and unstable manifolds of the hyperbolic periodic orbits, which bifurcate when the energy increases. The methodology presented here applies to other open systems with a double-well potential.

\end{abstract}

\maketitle


\section{Introduction}

During the last two decades, experimental \cite{Suits2020} and theoretical advances \cite{Bowman2017, Mauguiere2017} have established a new chemical reaction mechanism known as \emph{roaming}. In a conventional dissociation reaction, a molecule splits into two parts that immediately separate. In a roaming reaction, the molecule splits into two fragments. However, one part moves around the other until the two fragments collide, recombine, and finally, two new stable molecules separate. After pioneering work on formaldehyde (H$_2$CO) dissociation\cite{townsend2004roaming, Zhang2004}, roaming has been reported for other molecules \cite{Houston2006,Chao2011,Bowman2011roaming,Tsai2014, Tsai2015,Lombardi2016,BowmanRoaming,Mauguiere2017} and as well as for many bimolecular systems \cite{Christoffel2009,Takayanagi2011,Bencsura2012,Li2013,Joalland2013,Joalland2014,Hornung2015,Mauguiere2016,Cascarini2019}. Currently, roaming is a widely recognised chemical reaction mechanism.

Roaming reactions are impossible to explain with the conventional transition state theory based on configuration space. Transition state theory postulates chemical reactions as crossing events associated with a saddle point barrier separating reactants from products in a potential energy surface. However, roaming reactions are \emph{barrierless} \cite{Bowman2011roaming}. Previous work has demonstrated that information about the potential energy surface is insufficient for understanding roaming reactions and that motion of the dissociating molecules is essential \cite{Zhang2004,Lahankar2008, Bowman2017}. In this context, phase space provides a complete dynamical picture of the molecule's motion, rather than the limited picture obtained from the configuration space perspective alone. Consequently, recent efforts aim to extend transition state theory to a phase space description consistent with roaming. For a review on this topic, we refer to \cite{Mauguiere2017}.

A simple description of roaming in phase space is obtained from the study of classical trajectories of a dissociating diatomic molecule on a \emph{cirque} potential energy surface. Cirque potentials are a class of potentials that are almost flat, except for one or more localised depressions \cite{Quapp2007,Kraka2010}. Roaming pathways have been previously investigated in two-dimensional Hamiltonian models using cirque potentials constructed from overlapping two symmetric, and identical Morse potential wells \cite{Carpenter2017,Carpenter2018,GonzalezMontoya2020}. In this context, a roaming dissociation is described by trajectories that leave one well and enter the other without crossing the saddle energy barrier in between. These trajectories move between the wells by making excursions over the almost flat region surrounding the two wells. This model found that phase space objects guide roaming trajectories. In general, hyperbolic (or unstable) periodic orbits play an essential role in organising phase space transport in Hamiltonian systems with two or more degrees of freedom \cite{Wiggins2001,Wiggins08,wig2016}. Their stable and unstable manifolds govern the lobe dynamics in phase space, and they are essential for understanding the roaming reaction mechanism \cite{Mauguiere2017}. For a detailed description of how phase space structures are central for the study of chemical reaction dynamics, see \cite{agaoglou2019}.

This work introduces a new cirque potential energy surface by combining two van der Waals potentials and investigating the roaming dynamics in the resulting 2-degree of freedom Hamiltonian system. In contrast to a Morse potential, a van der Waals potential models a weak force between two neutral molecules via the forces obtained from the multipole expansion. We study a simple 2-dimensional restricted van der Waals model as an initial step for understanding the dynamics in 3 and more dimensions in the same spirit that the restricted planar 3-body problem in celestial mechanics provides an understanding of the restricted 3-body problem. We focus on two different reaction mechanisms: inter-well transport and transport between either well and the almost flat asymptotic region. We do so for an energy regime above the dissociation energy, where the hyperbolic periodic orbits exist. Their 2-dimensional stable and unstable manifolds play an essential role in the transport in each constant energy manifold.

The outline of the present paper is as follows. Section \ref{sec:potential} describes the basic features of the cirque potential energy surface considered. Section \ref{sec:lagrangian_descriptors} introduces the method of Lagrangian descriptors based on the action integral that we use to visualise the stable and unstable manifolds of hyperbolic periodic orbits that characterise the transport in phase space. We also apply this technique to find the bifurcations of hyperbolic periodic orbits as the energy of the system increases. Section \ref{sec:results} discusses the results of our analysis employing Lagrangian descriptor plots and fate maps determining the different transport routes that exist in this system. Finally, in Section \ref{sec:conclusion} we present the conclusions and remarks on this work.

\section{The Cirque Potential Energy Surface Model}
\label{sec:potential}

In this section, we construct a cirque double-well potential energy, $V(x, y)$, by combining two van der Waals potentials $W(x,y)$. This approach is similar to the one carried out for the double Morse potential in \cite{Carpenter2017, Carpenter2018,GonzalezMontoya2020}. The van der Waals potential $W(x,y)$ has been proposed to study the escape time in an ultracold chemical reaction in \cite{Soley2018}, and it has some similarities with the Morse potential well. Both potentials decay to zero asymptotically and have an unstable periodic orbit associated with the maximum of the effective potential. The potential energy surface model is obtained from the addition of two identical van der Waals potential energy surfaces, $W_1(x,y)$ and $W_2(x,y)$ centred at $(-d, 0)$ and $(d,0)$ respectively, with $d$ being the common distance away from the origin:

\begin{equation}
V(x,y) = W_1(x, y) + W_2(x, y)
= \dfrac{- W_0 \, k^6}{\left(\left(x + d\right)^2 + y^2 + k^2\right)^3} + \dfrac{- W_0 \, k^6}{\left(\left(x - d\right)^2 + y^2 + k^2\right)^3} \;.
\label{eq:double_vdw}
\end{equation}

\begin{figure}[htbp]
\centering
a)\includegraphics[scale=0.32]{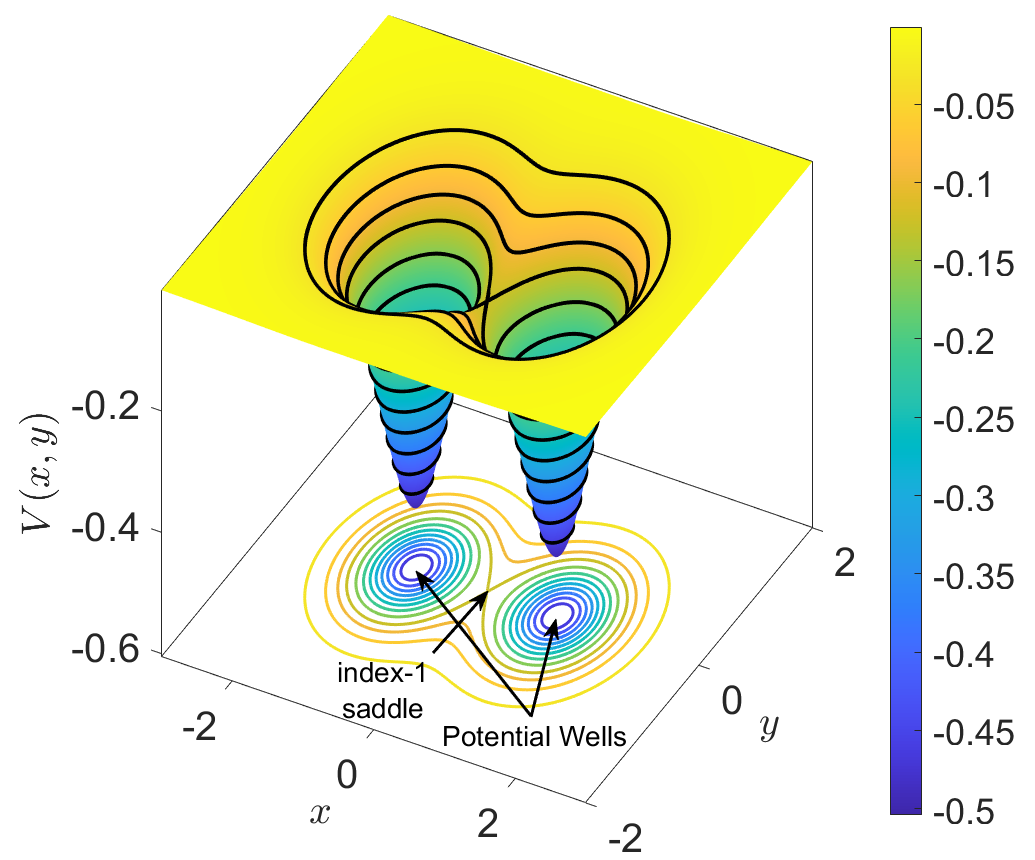}
b)\includegraphics[scale=0.31]{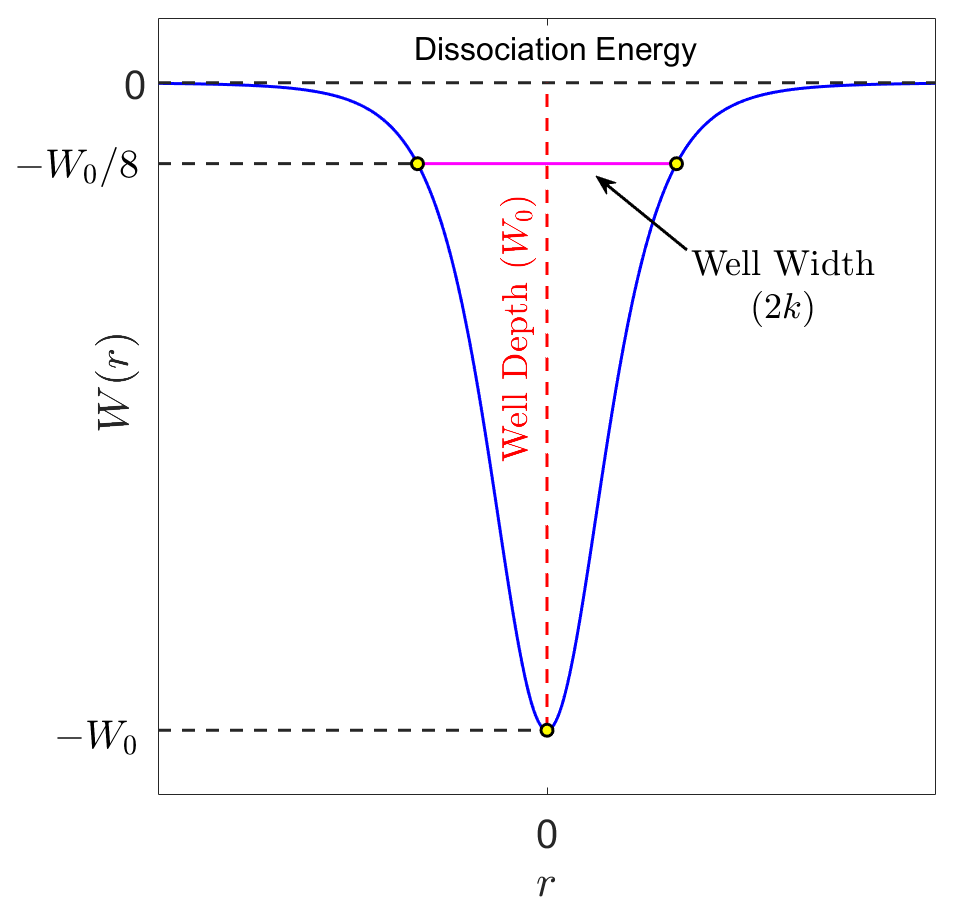}
\caption{(a) Double van der Waals potential for the parameters: $W_0 = 1/2$, $k = 1$, $d = 1$. (b) Geometrical interpretation of the parameters $W_0$ and $k$ for each potential well in Eq. \eqref{eq:double_vdw}.}
\label{fig:double_vdw_PES}
\end{figure}

Fig.\ref{fig:double_vdw_PES}(a) below shows a representation of the energy landscape for this potential energy surface. We observe the symmetry of $V(x,y)$ for the axes $x$ and $y$, a saddle point located at the origin between both wells, and that the potential energy surface becomes flat at infinity. The expression used for the van der Waals potential corresponding to each potential well, $W(x,y)$, is equivalent to the one proposed in \cite{Soley2018,Soley_thesis} to estimate the escape time in a cold chemical reaction. However, the parameters $W_0$ and $k$ used here provide a geometrical interpretation regarding the shape of $W(x,y)$. Figure  \ref{fig:double_vdw_PES}(b) illustrates how $W_0$ is equal to the well depth, while $k$ modulates the well width.  


\newpage

From the definition of the potential energy $V(x,y)$, it is clear that this potential is always negative. The phase space is bounded for $E<0$ and unbounded in any direction in the configuration space for $E>0$. The dissociation threshold energy is $E=0$. This work focuses on studying the dynamics for $E>0$ where the trajectories can escape to infinity.

An important difference between the double van der Waals potential energy and the double Morse potential is their different shapes in the centres of the wells. The double Morse has repulsive barriers corresponding to a forbidden region in the phase space that does not exist for the double van der Waals potential energy. Hence their phase spaces around these regions are very different. Section \ref{sec:results} discusses more details about these differences.

The 2 degrees-of-freedom Hamiltonian for this double van der Waals potential energy surface is defined as the sum of kinetic plus potential energy:
\begin{equation}
H(x,y,p_x,p_y) = \dfrac{p_x^2}{2 m} + \dfrac{p_y^2}{2 m} + V(x,y),
\label{eq:hamil}
\end{equation}

\noindent
where $m$ is the mass of the particle, $(x,y)$ are the configurations space coordinates, and $(p_x,p_y)$ are their conjugate momenta. The evolution of the dynamical system is determined by Hamilton's equations of motion:
\begin{equation}
\begin{cases}
\dot{x} = \dfrac{\partial H}{\partial p_x} = \dfrac{p_x}{m}  \\[.5cm]
\dot{y} = \dfrac{\partial H}{\partial p_y} = \dfrac{p_y}{m}  \\[.3cm]
\dot{p}_x = - \dfrac{\partial H}{\partial x} = - 6 W_0 \, k^6 \left[\dfrac{x - d}{\left(\left(x - d\right)^2 + y^2 + k^2\right)^4} + \dfrac{x + d}{\left(\left(x + d\right)^2 + y^2 + k^2\right)^4}\right] \\[.7cm]
\dot{p}_y = - \dfrac{\partial H}{\partial y} = - 6 W_0 \, k^6  \left[\dfrac{y}{\left(\left(x - d\right)^2 + y^2 + k^2\right)^4} + \dfrac{y}{\left(\left(x + d\right)^2 + y^2 + k^2\right)^4}\right]
\end{cases}
\label{eq:ham_eq}
\end{equation}

\noindent
Moreover, since energy is conserved, motion occurs in a 3-dimensional energy hypersurface embedded in the 4-dimensional phase space.

\section{Lagrangian Descriptors based on the Classical Action}
\label{sec:lagrangian_descriptors}

The mathematical technique used in this work to explore the phase space structures characterising the Hamiltonian system dynamics is the method of Lagrangian descriptors. This scalar visualisation tool was first introduced for the analysis of transport and mixing processes in the context of fluid dynamics \cite{madrid2009,mancho2013lagrangian,mendoza2010}. Its original intuitive formulation was given in terms of the computation of the arclength of trajectories generated from a grid of initial conditions of the dynamical system as they are integrated forward and backwards for a given time interval. The general idea behind this tool is to integrate a positive scalar quantity along the trajectories of the dynamical system. For example, other definitions of Lagrangian descriptors include the use of the $p$-norm of the components of the vector field that defines the flow of the system \cite{lopesino2017}, or the computation of the action integral \cite{montoya2020phase}, which is the approach we follow in this paper. For a comprehensive review on the Lagrangian descriptor method that includes an introductory tutorial on its application to classical Hamiltonian systems with a Python software package, we refer to \cite{ldbook2020}.

In the context of chemical reaction dynamics in Transition State Theory, this tool is a valuable technique for computing the phase space structures that govern the reactivity of the system   \cite{craven2015lagrangian,craven2016deconstructing,craven2017lagrangian,revuelta2019unveiling,agaoglou2019,Main2017,Main2018,Main2019,Bartsch2016}. The scaffolding for the determination of reaction rates is provided by the dividing surface, separating reactants from products, that is attached to an unstable periodic orbit. A crucial ingredient of the reaction rate is the flux of trajectories that cross the dividing surface as they evolve from one side of the transition state to the other. Along this path, they follow the stable and unstable manifolds of the hyperbolic periodic orbits, which act as a transport network of tube-like structures, guiding the evolution of the trajectories in the phase space. Other applications of Lagrangian descriptors to the study of chemical problems include the analysis of isomerisation reactions \cite{GG2020b,naik2020} and roaming \cite{krajnak2019,GonzalezMontoya2020}, the study of the influence of bifurcations on the manifolds that control chemical reactions \cite{GG2020a}. And also in the explanation of the dynamical matching mechanism in terms of the existence of heteroclinic connections in a Hamiltonian system defined by Caldera potential energy surface \cite{katsanikas2020a}.

In a Hamiltonian system where the Hamiltonian function has the form:
\begin{equation}
H(\mathbf{q},\mathbf{p}) = T(\mathbf{p}) + V(\mathbf{q}) \, ,
\label{class_ham}
\end{equation}

\noindent
where the kinetic energy $T$ is a quadratic function of the momenta $\mathbf{p}$, and the potential energy surface $V$ depends only on the configuration coordinates $\mathbf{q}$, we can construct a Lagrangian descriptor indicator by means of the classical action (Maupertuis action) $S_0$ calculated along a trajectory $\mathbf{x}(t) =  \left(\mathbf{q}(t),\mathbf{p}(t)\right)$ in the phase space of the system:
\begin{equation}
S_0\left[\mathbf{x}(t)\right] = \int^{\mathbf{q}_f}_{\mathbf{q}_i} \mathbf{p} \cdot d\mathbf{q}  \, ,
\label{action_def}
\end{equation}

\noindent
where $\mathbf{q}_i = \mathbf{q}(t_0)$ and $\mathbf{q}_f = \mathbf{q}(t_1)$ are the initial and final configuration space points on the trajectory. The action $S_0$ defines a natural metric in the phase space, which we use to construct a Lagrangian descriptor. Notice that we can rewrite the action integral as follows:
\begin{equation}
S_0\left[\mathbf{x}(t)\right] = \int^{\mathbf{q}_f}_{\mathbf{q}_i} \mathbf{p} \cdot d\mathbf{q} = \int^{t_1}_{t_0} \mathbf{p} \cdot \dfrac{d\mathbf{q}}{d t} \, dt = \int^{t_1}_{t_0} \dfrac{|\mathbf{p}|^2}{m} \, dt = 2 \int^{t_1}_{t_0}  \, T \, dt \, .
\label{action_ke}
\end{equation}

\noindent
in which case, we have assumed that the mass in each degree of freedom is the same, $m$.

Given any initial condition $\mathbf{x}_0 = \mathbf{x}(t_0)$, we construct its trajectory by evolving the initial condition forward and backward in time for times $\tau_{+}$ and $\tau_{-}$ respectively. The action-based Lagrangian descriptor evaluated along trajectories is defined as:
\begin{equation}
M_{S_0} (\mathbf{x}_{0},t_{0},\tau_{+},\tau_{-}) = M^{+}_{S_{0}}(\mathbf{x}_{0},t_{0},\tau_{+}) + M^{-}_{S_0}(\mathbf{x}_{0},t_{0},\tau_{-})
\label{eq:LD_S}
\end{equation}

\noindent
where $M^{+}_{S_{0}}$ and $ M^{-}_{S_0}$  correspond, respectively, to the forward and backward contributions of the action integral to the Lagrangian descriptor scalar field:
\begin{equation}
 M^{+}_{S_{0}}(\mathbf{x}_{0},t_{0},\tau_{+}) = \int^{ \mathbf{q}_{+}}_{ \mathbf{q}_{0}} \mathbf{p} \cdot d\mathbf{q} = 2 \int^{t_{0}+\tau_{+}}_{t_{0}} T \; dt \quad,\quad M^{-}_{S_0}(\mathbf{x}_{0},t_{0},\tau_{-}) =  \int^{\mathbf{q}_{0}}_{\mathbf{q}_{-}} \mathbf{p} \cdot d\mathbf{q} = 2 \int^{t_{0}}_{t_{0}-\tau_{-}} T \; dt \,.
\label{eq:LD_S_fwbw}
\end{equation}

When this scalar function is applied to a given set of initial conditions in the phase space, it produces a scalar field that has the capability of highlighting the location and geometry of the invariant stable and unstable manifolds associated with the normally hyperbolic invariant manifolds (NHIMs) \cite{wig2016,naik2019a}. These invariant manifolds can be easily identified with features of the Lagrangian descriptor output, where the scalar field displays abrupt changes in its values. It is important to remark that if we consider only the contribution from the forward integration of trajectories, this will provide information about the stable manifolds. At the same time, the backward term is used to determine the unstable manifolds. Suppose both terms of Lagrangian descriptors are added. In that case, the output can be used to visually locate the presence of NHIMs in the system at the intersection of the stable and unstable manifolds. In the case of a 2-degree of freedom Hamiltonian system, unstable periodic orbits are NHIMs.


The abrupt changes in the scalar fields are generated by the different behaviour of the trajectories nearby. The trajectories trapped in the stable manifold converge to a periodic orbit. Meanwhile, the trajectories around the stable manifold have very different behaviour for long times. Consequently, their corresponding values of action grow differently. For this reason, numerically, the Lagrangian descriptor evaluated on a finite number of points has a jump in the intersection of the set of initial conditions with the stable and unstable manifolds and in the boundaries of KAM islands.

An important difference between the Lagrangian descriptors and the classical escape time for open systems is that the trajectory integration is not only necessarily stopped when the particle reaches the asymptotic region for the Lagrangian descriptor. The integration time for the Lagrangian descriptor $\tau_+$ and $\tau_-$ can be chosen freely. A practical choice in this work is stopping the integration when the particle reaches a specific region of the configuration space, then some of the abrupt changes in the Lagrangian descriptor plots are related to the stopping of the integration. The singularities in the escape time correspond to trapped trajectories. The neighbouring trajectories spend more time when they are closer to the stable manifolds. In the limit, when the initial condition is chosen closer to a stable manifold, the escape time goes to infinity.
 
Another exciting aspect of the Lagrangian descriptor based on the action $S_0$ is its interpretation based on the values of the potential energy $V(\mathbf{q})$ behaviour of the trajectories $\mathbf{q(t)}$ \cite{montoya2020phase}. If a trajectory spends a considerable time in a region with large values of $V(\mathbf{q})$, then its Lagrangian descriptor is less than a trajectory that spent the same time in a region with lower values of $V(\mathbf{q})$. This fact gives us information about the trajectories from its Lagrangian descriptor values, and it is beneficial for analysing open Hamiltonian systems.
 

\section{Transport Routes and Roaming}
\label{sec:results}

In this section, we study the phase space transport associated with the cirque potential energy surface. For the numerical example, the values of the parameters of the potential energy surface are $W_0 = 1/2$, $k = 1$ and $d = 1$. The potential energy surface $V(x,y)$ shares basic common features with other potentials where roaming has been observed. The potential energy has two wells surrounded by a region that becomes almost flat as the configuration space coordinates asymptotically approach infinity, see Fig.\ref{fig:double_vdw_PES}. The dissociation energy threshold for the system is $E = 0$, which corresponds to the critical value where the phase space of the system goes from bounded, $E < 0$, to unbounded, $E > 0$. Then, for positive energy values, the trajectories can escape from the potential wells to infinity. In some regions, the motion can be complicated, like a chaotic motion only for some finite time interval before escape to the asymptotic region. This phenomenon is called transient chaos, and it is characteristic in open systems \cite{Tel_book,Sanjuan2013,Tel2015}, some recent examples of 3-degree of freedom Hamiltonian has been studied in \cite{Drotos2014,Gonzalez2014,Gonzalez2020_atom,Waalkens2020,Agaoglou2021}.

\begin{figure}[htbp]
(a)\includegraphics[scale=0.35]{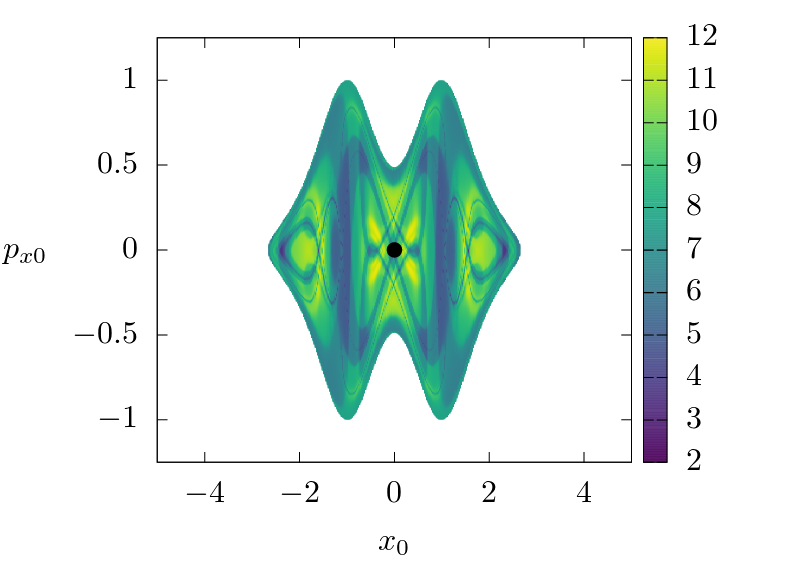}
(b)\includegraphics[scale=0.35]{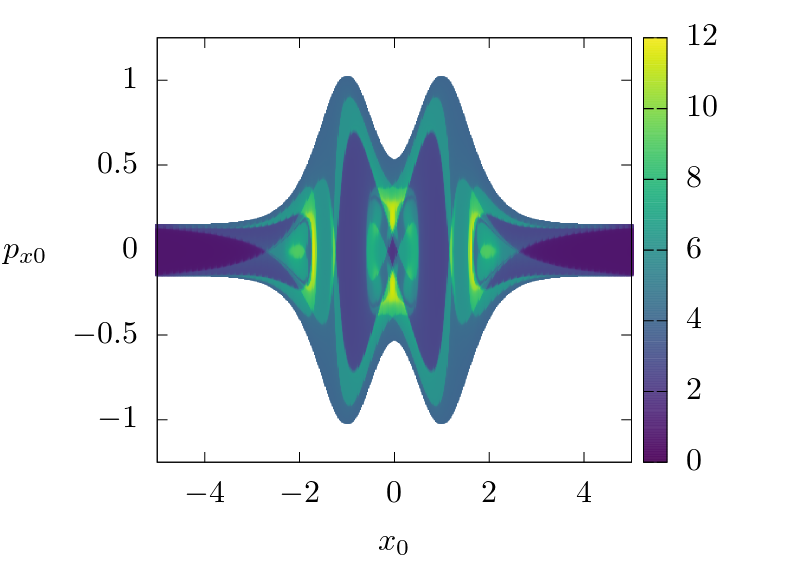}
(c)\includegraphics[scale=0.35]{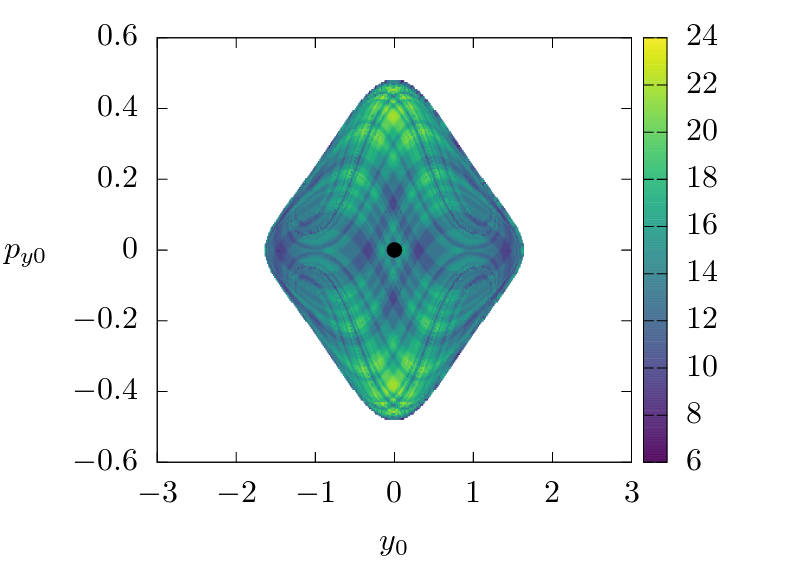}
(d)\includegraphics[scale=0.35]{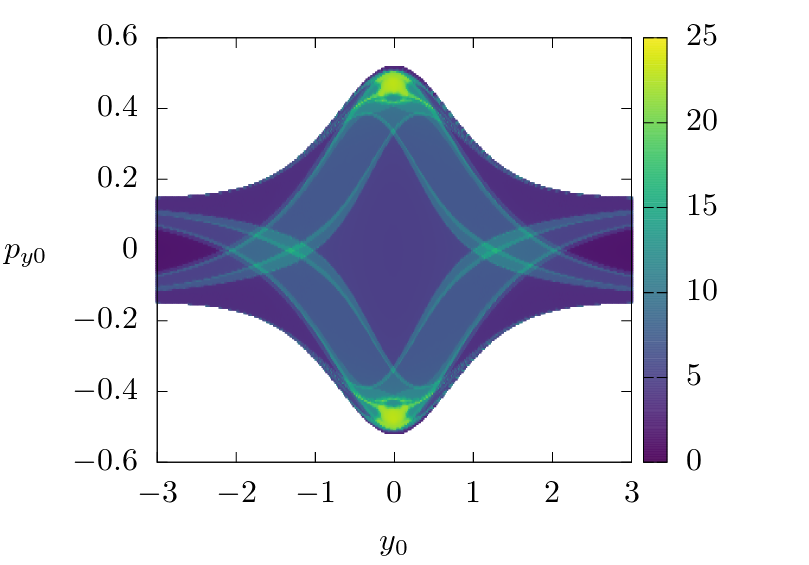}
\caption{Lagrangian descriptor $M_{S_0}$ for $E =-0.01$ (a) and (c) and $E = 0.01$ (b) and (d) on the canonical conjugate planes $x_0$-$p_{x0}$ for $y=0$ (a) and (b), and $y_0$-$p_{y0}$ for $x = 0$ (c) and (d). Integration times are $\tau = 20$ and $\tau = 40$, respectively. Momenta are sampled according to the energy conservation condition. For negative energies, the phase space is bounded and for positive ones it is unbounded, and thus trajectories can escape to infinity. In panels (b) and (d) the hyperbolic periodic orbit $\Gamma_1$ that surrounds both wells is visible, its stable and unstable manifolds intersect the plane around the points $(x_0,p_{x0}) = (2.6,0)$ and $(y_0,p_{y0}) = (2.1,0)$. The periodic orbits contained in the symmetry axis $x$ and $y$ disappear for $E>0$, compare panels (a) and (c) with (b) and (d).
}
\label{fig:ld_x_px}
\end{figure}

The Lagrangian descriptor plots in Fig.\ref{fig:ld_x_px} show how the phase space structures change as the energy of the system goes from negative, $E = -0.01$, to positive, $E = 0.01$. The set of initial conditions for these calculations are the canonical planes $x$--$p_x$ and $y$--$p_y$. Notice that the structure of the tangles formed between the stable and unstable manifolds changes considerably. There are two important bifurcations of periodic orbits that occur at the critical energy $E = 0$. In the first bifurcation, the two unstable periodic orbits along the potential energy symmetry lines $x$ and $y$ disappear. The system's phase space becomes unbounded at this energy, and the turning points of those periodic orbits disappear at infinity. In the phase space, the lines $x,p_x=0$ and $y,p_y=0$ are invariant manifold due to the discrete symmetries of the vector field that define the system. However, they are not hyperbolic. There are no directions that attract or repels at infinity, and the particles behave as free particles in the asymptotic region \cite{Newton_book}. The potential of Eq.\ref{eq:double_vdw} falls like $r^{-6}$ for large distances. This behaviour is sufficient to fulfil the asymptotic conditions of classical scattering, and asymptotic trajectories are just the straight lines of free motion with constant velocity. The black dots in panels (a) and (c) Fig.\ref{fig:ld_x_px} indicate the intersection of the orbits and the set of initial conditions. A hyperbolic periodic orbit surrounding the region of both potential wells, $\Gamma_1$, is created in the second bifurcation. This orbit plays an important role in the transport for $E>0$. The intersections of $\Gamma_1$ with the plane $x$--$p_x$ corresponds to a point with an abrupt jump in the Lagrangian descriptor with largest values of $x_0$ on the line $p_{x0} = 0$ in panel (b) of Fig.\ref{fig:ld_x_px}. Similarly, the intersections of $\Gamma_1$ with the plane $y$--$p_y$ corresponds to points with an abrupt jump in the Lagrangian descriptor with largest values of $y_0$ on the line $p_{y0}$ in panel (d) of Fig.\ref{fig:ld_x_px}.

A trajectory of the system is classified as a roaming trajectory whenever it evolves from one potential well to the other through visiting the region where the potential energy surface is almost flat, and it does so without crossing the region close to the saddle point between both wells. This behaviour is associated with a particular chemical reaction where a molecule breaks up into two parts. One part of the molecule orbits around the other for some time before both parts recombine to break up and form new molecules. This kind of chemical reaction is impossible to explain if the motion of the individual fragments of the molecule is not considered. Hence, a roaming reaction occurs only for a unique combination of momentum and position of the reactants.

\begin{figure}[htbp]
\centering
\includegraphics[scale=0.9]{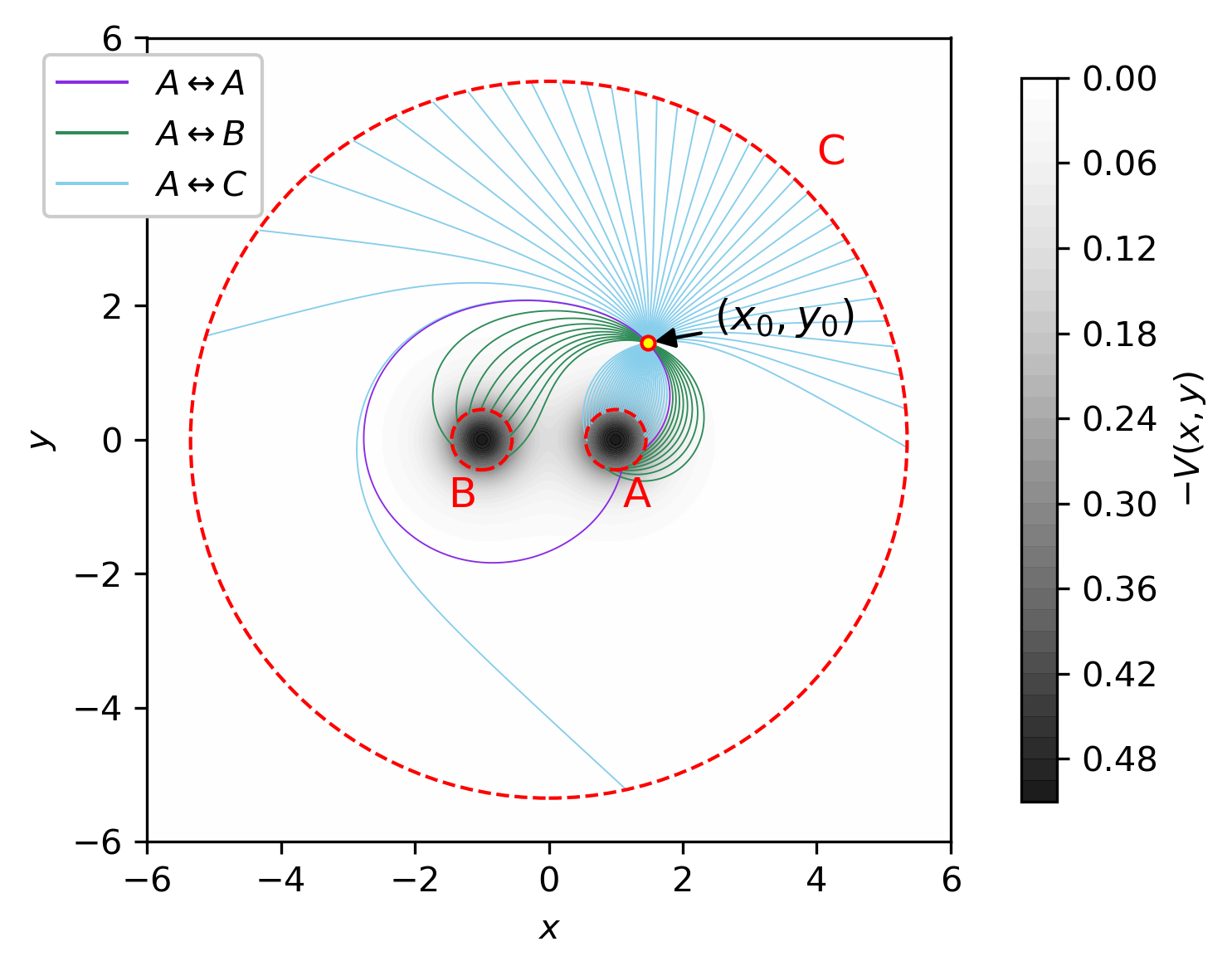}
\caption{Transport between the three configuration space regions $A$, $B$, and $C$. In this example, trajectories start their evolution in region $A$ and travel through the region where the potential energy surface is almost flat, ending up in regions $A$, $B$, or $C$. The trajectories from $A$ to $B$, $A$ to $C$ , and $A$ to $A$ are in colours green, skyblue, and violet, respectively. All the trajectories cross the point $(x_0,y_0)$ in the configuration space. The potential energy $V(x,y)$ is in grey scale in the background.}
\label{fig:regions_A_B_C_potencial}
\end{figure}

Following a similar approach as that carried out in previous studies of roaming for the double Morse potential well \cite{Carpenter2017, Carpenter2018, GonzalezMontoya2020}, we consider for the analysis three regions in configuration space: regions $A$ and $B$ enclosing the minimum of each well are defined by two circles centred at $(x,y) = (0,\pm d)$, respectively, with an equal radius $\sim 0.5$. On the other hand, region $C$, which corresponds to the asymptotic region, is delimited by a circle centred at the origin and with radius $\sim 5.5$ containing the two other regions $A$ and $B$, see Fig \ref{fig:regions_A_B_C_potencial}. Trajectories in the region $C$ behave as almost-free particles since the potential energy surface gradient is close to zero in the asymptotic region. The boundaries of the regions were chosen as circles for simplicity.

Now, let us consider the distinct dynamical fates of trajectories that travel from:
\begin{itemize}
\item one potential well to the other ($A \rightarrow B$ and $B \rightarrow A$).
\item one well to the same well ($A \rightarrow A$ and $B \rightarrow B$).
\item one potential well to the asymptotic region ($A \rightarrow C$ and $B \rightarrow C$).
\item the asymptotic region to one well ($C \rightarrow A$ and $C \rightarrow B$).
\item the asymptotic region to itself ($C \rightarrow C$).
\end{itemize}

\begin{figure}[htbp]
\centering
\includegraphics[scale=0.88]{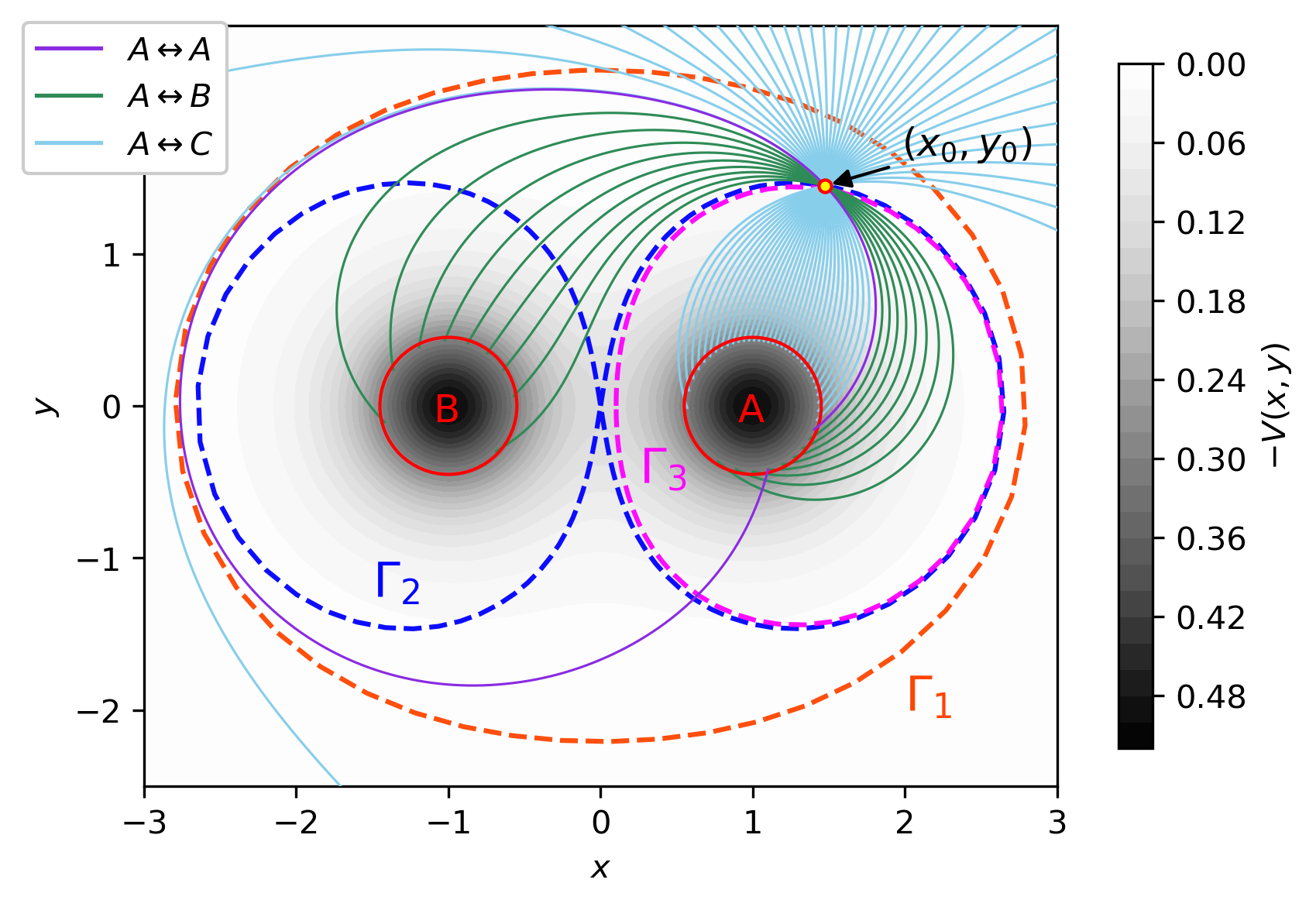}
\caption{Configuration space projection of the trajectories on the figure \ref{fig:regions_A_B_C_potencial} and the hyperbolic periodic orbits $\Gamma_1$, $\Gamma_2$, and $\Gamma_3$ associated with the transport mechanism for the energy $E$ = 0.01. The three orbits are in dotted lines. The potential energy $V(x,y)$ is in grey scale in the background. }
\label{fig:periodic_orbits_roaming}
\end{figure}

\begin{figure}[htbp]
\centering
\includegraphics[scale=0.34]{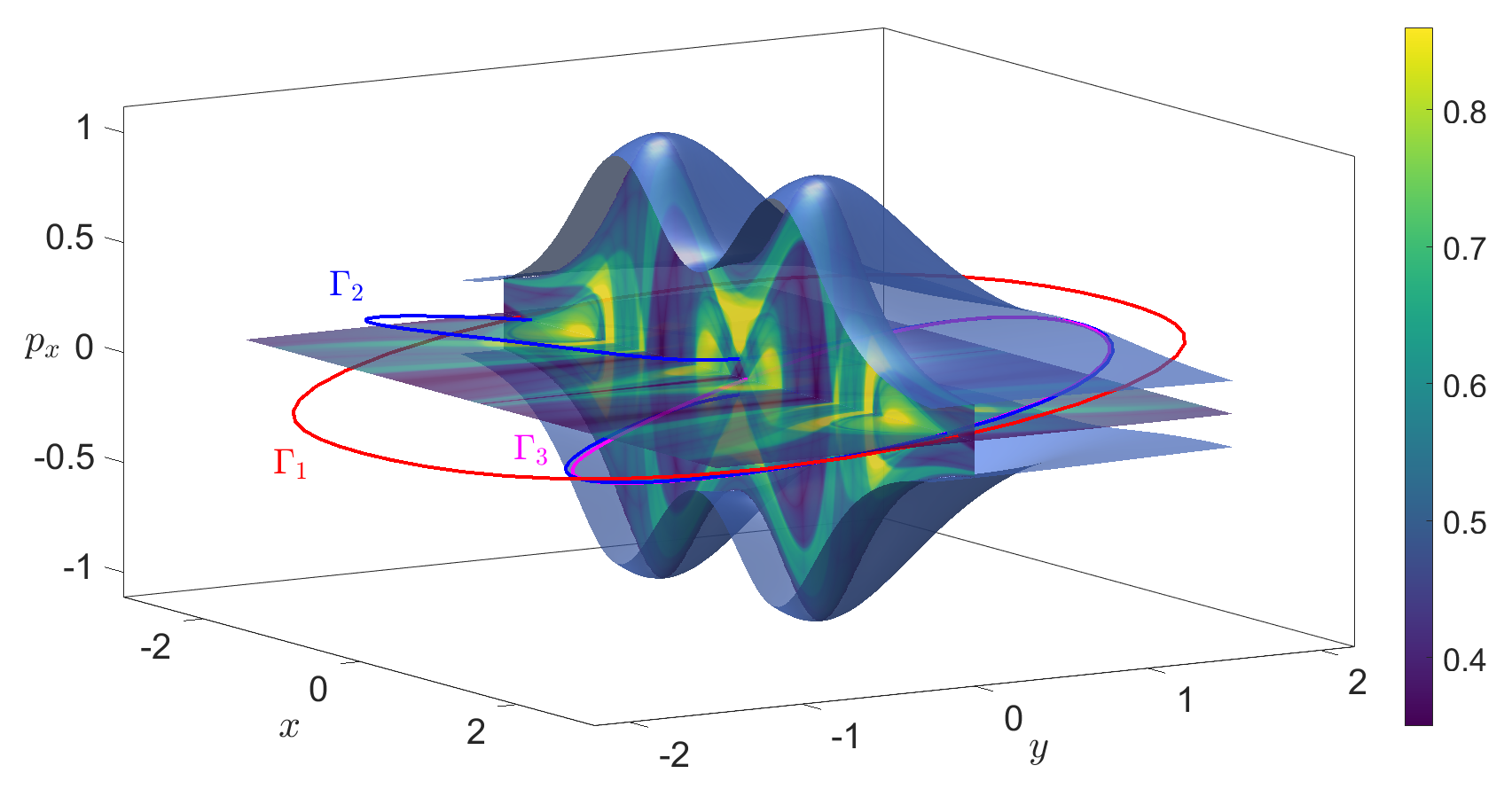}
\caption{Phase space representation of the hyperbolic periodic orbits $\Gamma_1$, $\Gamma_2$, and $\Gamma_3$ associated with the transport for the energy $E = 0.01$. The action based Lagrangian descriptors calculated for $\tau = 20$ on the planes $y = 0$ and $p_x = 0$ are depicted. Also, a section of the energy shell is included in metallic blue.}
\label{fig:periodic_orbits_roaming_phase_space}
\end{figure}

In the double van der Waals system, three families of unstable periodic orbits control the transport in phase space like in the double Morse system studied in \cite{Carpenter2017, Carpenter2018, GonzalezMontoya2020}. These unstable periodic orbits exist for energies higher than the threshold $E = 0$, and we can relate their projections onto configuration space with the geometry of the potential energy surface. The projection of the first family of periodic orbits, $\Gamma_1$, encircles both potential wells. The second family of periodic orbits, $\Gamma_2$, encircles both potential wells and crosses the origin. Finally, the projection of the unstable periodic orbits corresponding to the third family, $\Gamma_3$, encircle each potential well separately. See Figs \ref{fig:periodic_orbits_roaming} and \ref{fig:periodic_orbits_roaming_phase_space} for a visual representation of these hyperbolic periodic orbits in configuration space and their 3-dimensional phase space geometry inside the energy hypersurface.

Using Lagrangian descriptors and among other phase space structure indicators, we can find values of parameters of the system for which families of hyperbolic periodic orbits exist and bifurcate \cite{Demian2017,Gonzalez2012,GonzalezMontoya2020}. The main idea to find the bifurcation parameter values is to calculate the Lagrangian descriptor evaluated on a set of initial conditions that contain the intersection of the stable or unstable manifolds of the periodic orbits for a region of the parameter space. The dissipation of the singularities in the Lagrangian descriptor as a function of the parameters indicates the critical values of parameters such that the stable and unstable manifolds of the periodic orbits disappear, and the hyperbolic periodic orbits bifurcate.

\begin{figure}[htbp]
(a) \includegraphics[scale=0.39]{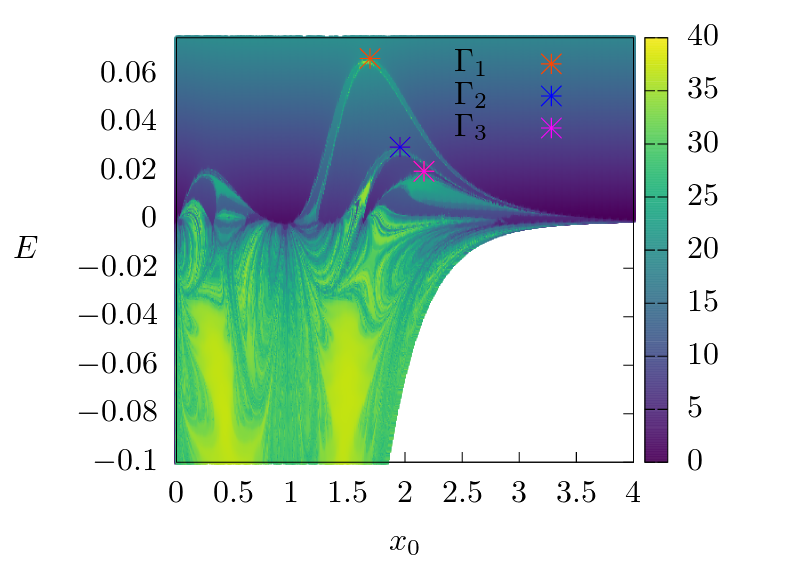}
(b)\includegraphics[scale=0.39]{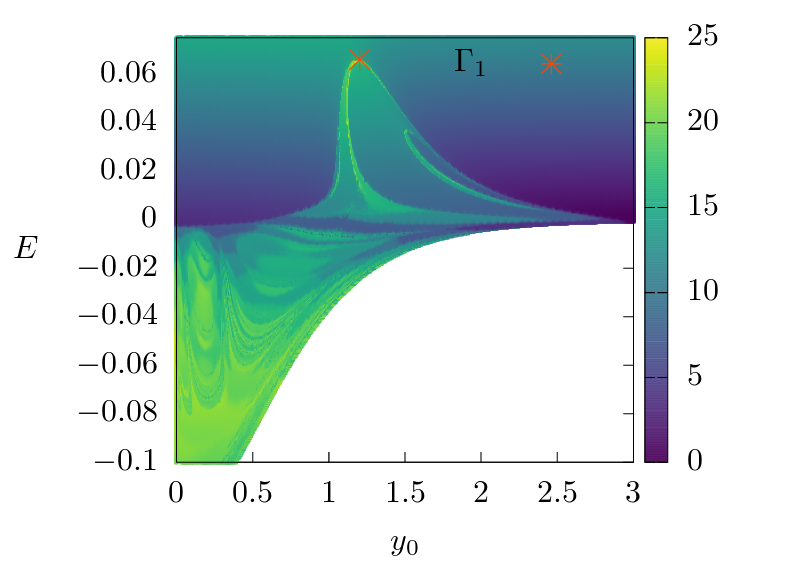}
\caption{Action-based Lagrangian descriptor $M^{+}_{S_0}$ as a function of the energy $E$ evaluated on the initial conditions $(x_0,y_0=0,p_{x0}=0,p_{y0}>0)$ and $(x_0=0,y_0,p_{x_0}>0,p_{y0}=0)$. The integration time is $\tau=60$. This type of Lagrangian descriptor plot is useful for finding the values of the parameters corresponding to abrupt changes in the phase space like bifurcation of hyperbolic periodic orbits. The periodic orbits $\Gamma_1$, $\Gamma_2$, and $\Gamma_3$ intersect the set of initial conditions where the Lagrangian descriptor is evaluated in panel (a). Meanwhile, only $\Gamma_1$ intersect the set of initial conditions corresponding to the panel (b). The starting points shows the bifurcation point in the diagram where the hyperbolic periodic orbits disappear.
}
\label{fig:ld_E_xy}
\end{figure}

In this study, we take $E$ as the bifurcation parameter. To find the values of $E$ for which the hyperbolic periodic orbits $\Gamma_1$, $\Gamma_2$, and $\Gamma_3$ disappear, we calculate a Lagrangian descriptor $M_{S_0}$ for a set of initial conditions, $(x_0, y_0, p_{x0}, p_{y0})$, that contains the periodic orbits. Each family of hyperbolic periodic orbits crosses the $x_0$-axis with momentum $(p_{x0} = 0, p_{y0} > 0)$. The Lagrangian descriptor corresponding to this set of initial conditions is shown in Fig.\ref{fig:ld_E_xy} (a) and the numerical values of the energies and intersection with the $x_0$ axis are in table \ref{table:bifurcations}. The starting points magenta ($\Gamma_3$), blue ($\Gamma_2$), and red ($\Gamma_1$) corresponds to the bifurcations.

\hspace{1.0cm}

\begin{table}
\centering
\begin{tabular}{c c c}
           &  $E$\hspace{0.3cm}  & $x_0$  \\
           \hline
$\Gamma_1$\hspace{0.3cm} & 0.066\hspace{0.3cm} & 2.144   \\
$\Gamma_2$\hspace{0.3cm} & 0.029\hspace{0.3cm} & 1.920   \\
$\Gamma_3$\hspace{0.3cm} & 0.019\hspace{0.3cm} & 1.678   \\

\end{tabular}

\hspace{1.0cm}

\caption{\label{table:bifurcations} Bifurcation energies for the periodic orbits $\Gamma_1$, $\Gamma_2$, and $\Gamma_3$ and their corresponding intersection with the axis $x_0$.}

\end{table}

Figure \ref{fig:ld_E_xy} (b) shows the Lagrangian descriptor evaluated on the line $x_0=0$ with momentum $(p_{x0} > 0, p_{y0} = 0)$. From the three hyperbolic periodic orbits, only $\Gamma_1$ intersects this set of initial conditions. The maximum around the point $(y_0,E)$ = $(1.186,0.066)$ corresponds to its bifurcation point. The value of the bifurcation energy is consistent for both sets of initial condition.

\begin{figure}[htbp]
(a) $\tau=20$ \includegraphics[width=0.375\textwidth]{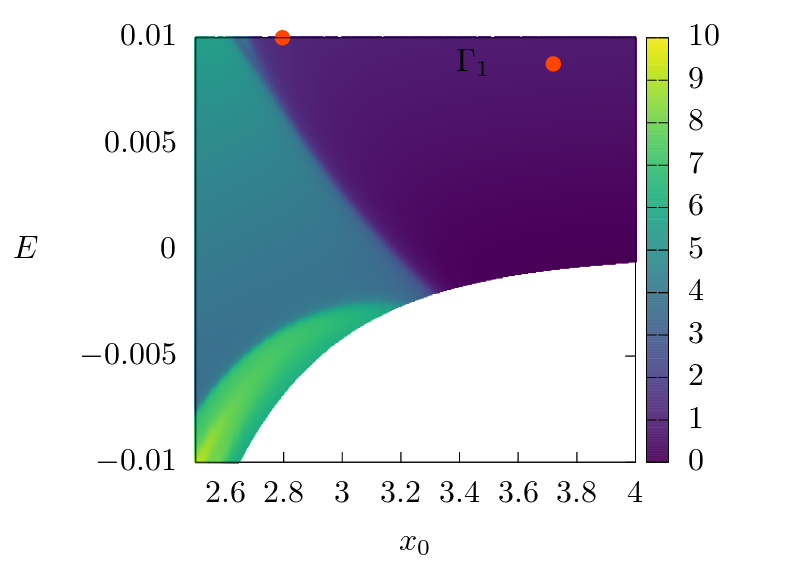}
(b) $\tau=40$\includegraphics[width=0.375\textwidth]{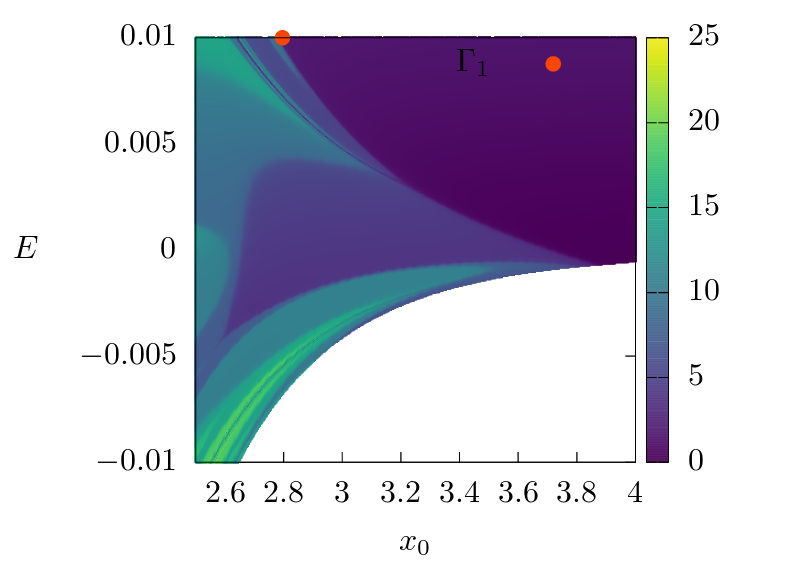}
(c) $\tau=60$\includegraphics[width=0.375\textwidth]{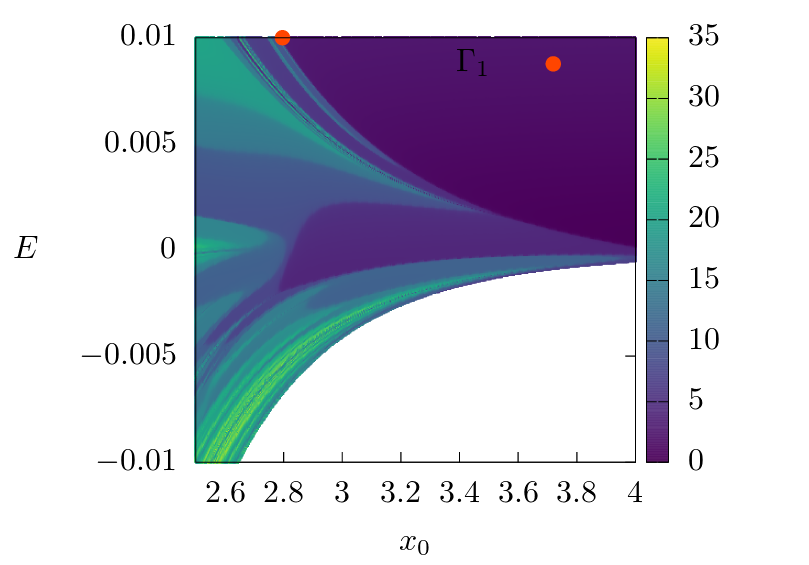}
(d) $\tau=200$\includegraphics[width=0.375\textwidth]{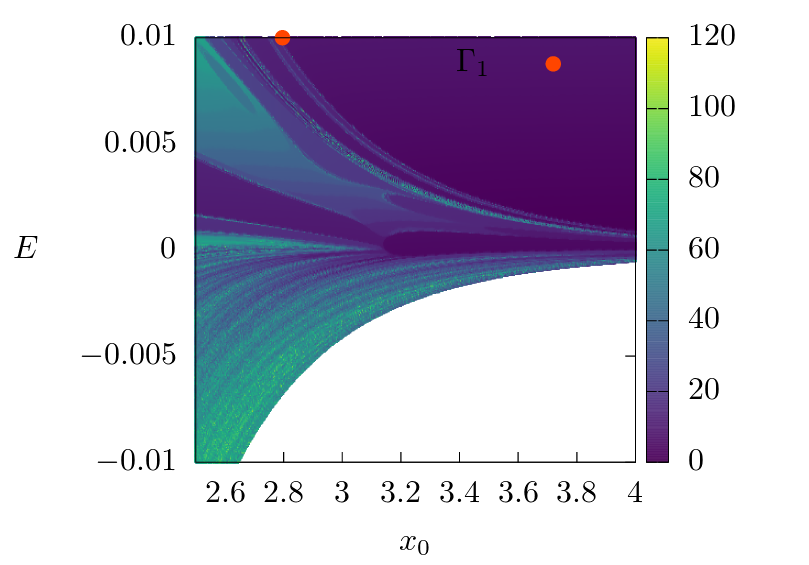}

\caption{Lagrangian descriptor $M^{+}_{S_0}$ as a function of the energy $E$ evaluated on the initial conditions $(x_0,y_0=0,p_{x0}=0,p_{y0}>0)$. The integration times are $\tau=20,40,60,200$. The plots show how the phase structures are revealed as the integration time increases. The red point is the intersection of the hyperbolic periodic orbit $\Gamma_1$ with the set of initial conditions for $E=0.01$. The jump in the Lagrangian descriptor plot for these value of the energy converges to the red intersection point when the integration time increases.  
  }
\label{fig:ld_E_x_zoom}
\end{figure}

Figure \ref{fig:ld_E_x_zoom} shows the convergence of the Lagrangian descriptor evaluated in a subdomain of Fig.\ref{fig:ld_E_xy} (a) when the integration time is increased. More intersections of stable manifolds with the set of initial conditions are visible for longer integration times. The line corresponding to the largest values of $x_0$ for each value of $E$ corresponds to the intersection of the stable manifold of the periodic orbit $\Gamma_1$.

In order to study transport between the wells and the roaming trajectories, we consider the dividing surfaces associated with each of the three periodic orbits $\Gamma_1$, $\Gamma_2$, and $\Gamma_3$. These surfaces have been previously used to study the double Morse potential energy surface in \cite{Carpenter2018, GonzalezMontoya2020}. The dividing surface constructed from a periodic orbit has two important properties for the analysis of the transport:

\begin{itemize}
    \item The periodic orbits associated with the dividing surface are subsets of the dividing surface.
    \item For any trajectory that crosses the dividing surface, its projection in configuration space crosses the periodic orbit projection.
\end{itemize}

The last property is essential for our study of the transport between potential wells and roaming. It allows us to visualise all the regions with different fates on the dividing surface. The algorithm to construct the dividing surface of a periodic orbit consist of 3 steps:

\begin{itemize}
    \item Project the periodic orbit on the configuration space.
    \item For each point $(x,y)$ on the configuration space, construct the circle compatible with energy conservation:
    \begin{equation}
    p^2_x + p^2_y =  2m(E - V(x,y)).
    \label{eq:circ_energy}
    \end{equation}
    \item Take the union of all the circles for any point in the projection of the periodic orbit.
\end{itemize}

An interesting question is: What is the topology of the dividing surfaces in the phase space? In the case of the hyperbolic orbit associated with a saddle point in the potential energy, the projection of the periodic orbit is a segment line in the configuration space which extreme points are in the equipotential set corresponding to the energy of the orbit, where the kinetic energy of the periodic orbit vanishes. The dividing surface for this kind of periodic orbit is a sphere that naturally divides the phase space into two sides around the neighbourhood of the saddle point in the potential energy \cite{Pechukas1973,Pollak1978,Pechukas1979,Waalkens2004}. If the periodic orbits are not associated with saddle points in the potential energy, their topology could be different. The projection of the periodic orbits $\Gamma_1$ and $\Gamma_3$ are closed curves that not self-intersecting, and then their diving surfaces are tori of genus one. Meanwhile, the projection of the periodic orbit $\Gamma_2$ is a curve with one intersection point, then the dividing surface of $\Gamma_2$ is a torus of genus 2.

\begin{figure}[htbp]
(a)\includegraphics[scale=0.39]{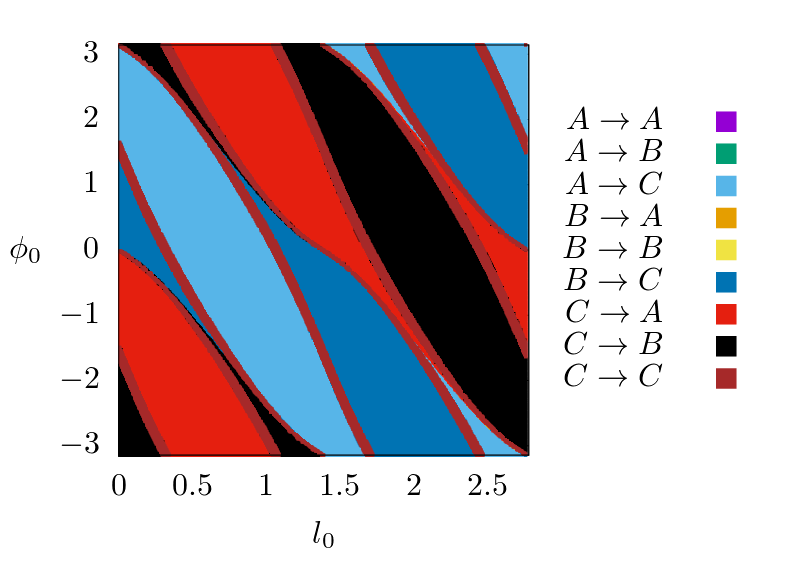}
(b)\includegraphics[scale=0.39]{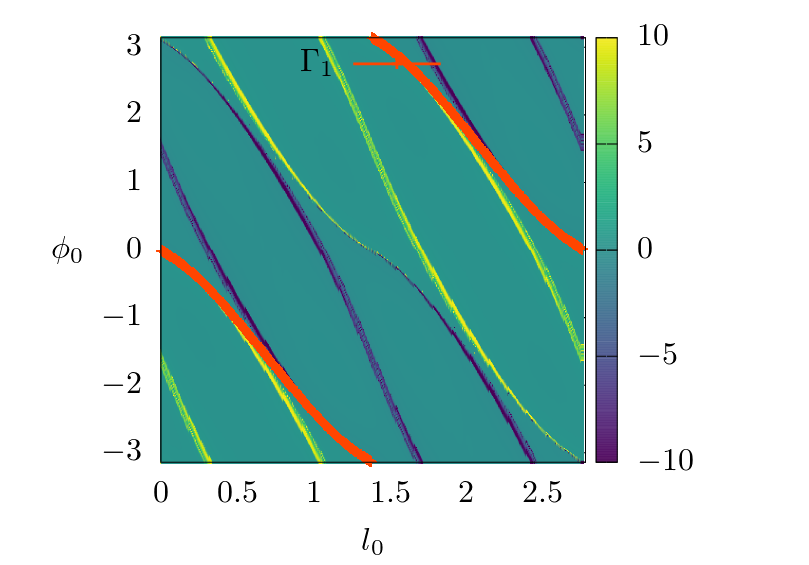}
(c)\includegraphics[scale=0.39]{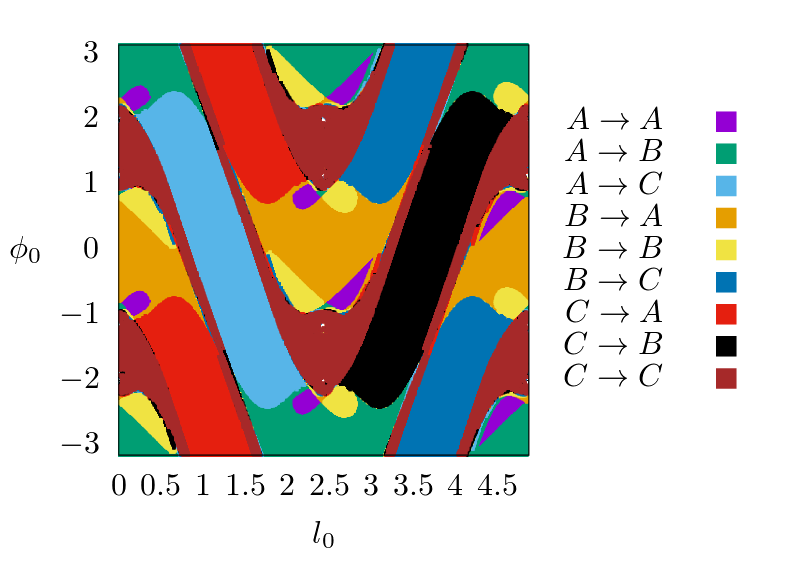}
(d)\includegraphics[scale=0.39]{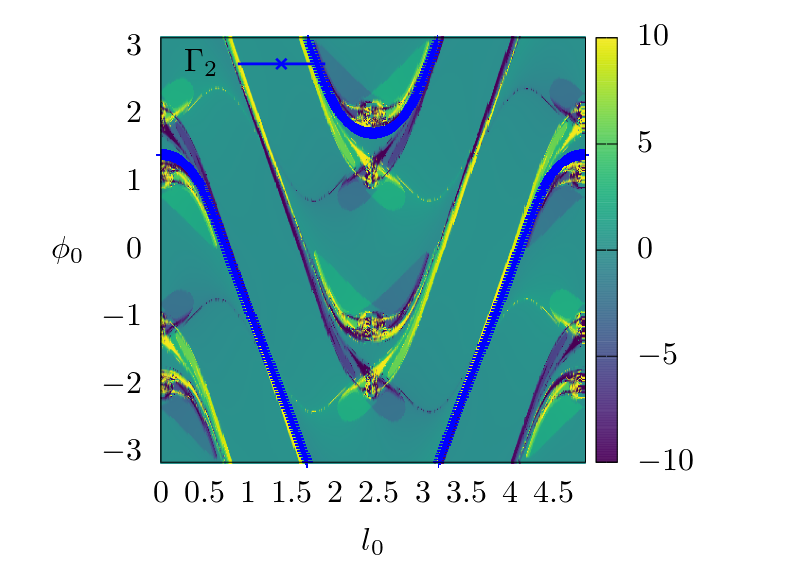}
(e)\includegraphics[scale=0.39]{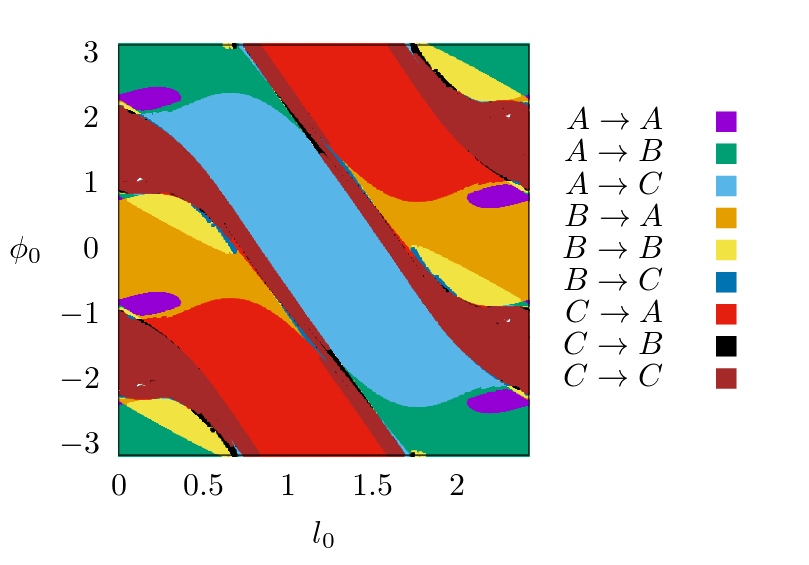}
(f)\includegraphics[scale=0.39]{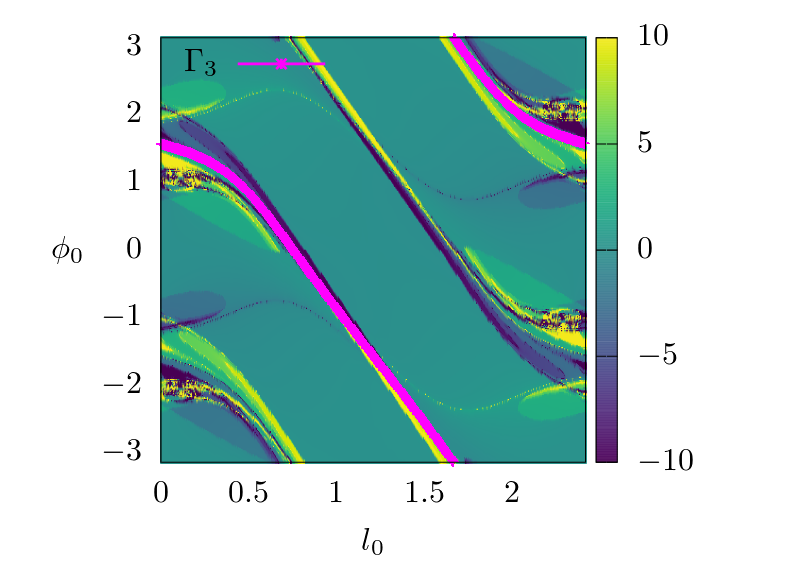}
\caption{Fate maps on the left and $M^{\pm}_{S_0}$ on the right evaluated on the dividing surface corresponding to periodic orbits $\Gamma_1$, $\Gamma_2$, and $\Gamma_3$.
The parts (a) and (b) correspond to the dividing surface of $\Gamma_1$. The parts (c) and (d) correspond to the dividing surface of $\Gamma_2$. The parts (e) and (f) correspond to the dividing surface of $\Gamma_3$. The energy of the systems $E$ = 0.01 is slightly above the dissociation threshold. The dividing surface is parametrised by the angle $\phi_0$ of the momentum for the $p_x$ component, and $l_0$ the arclength defined as the action integral $S_0$ along the periodic orbit.}
\label{fig:ld_fm_ds_E_001}
\end{figure}

\begin{figure}[htbp]
(a)\includegraphics[scale=0.39]{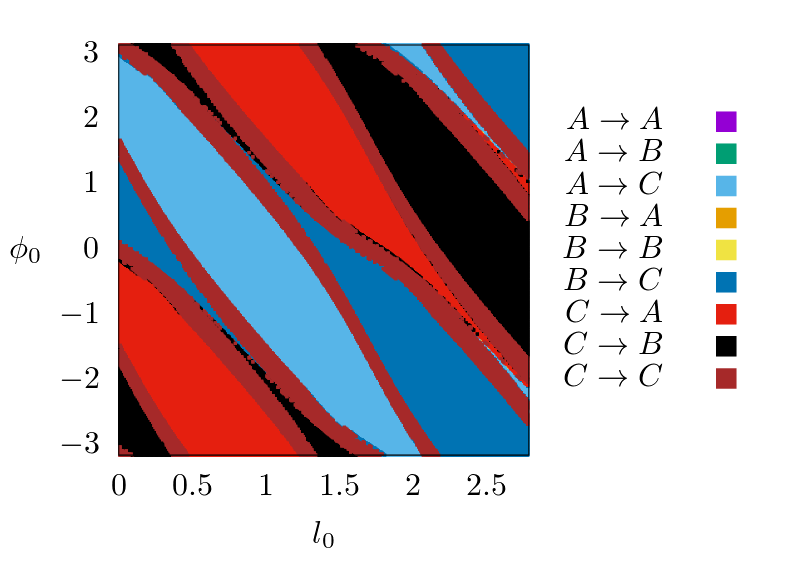}
(b)\includegraphics[scale=0.39]{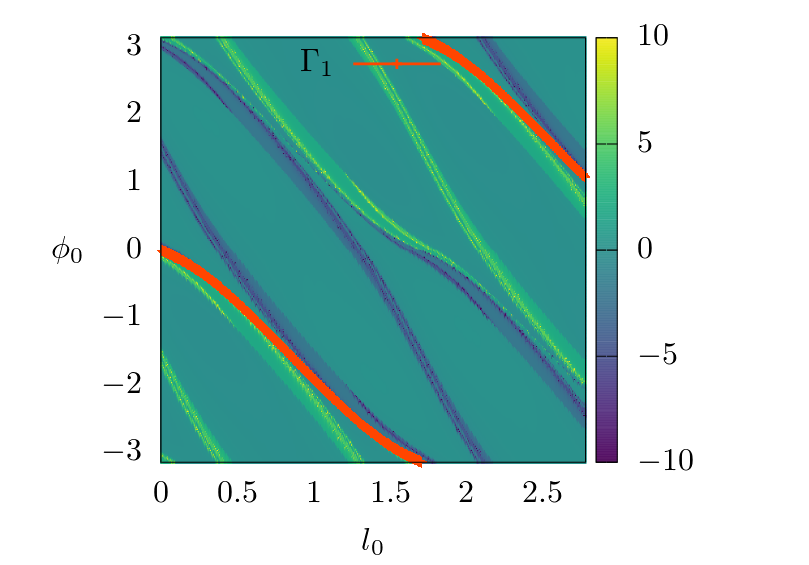}
(c)\includegraphics[scale=0.39]{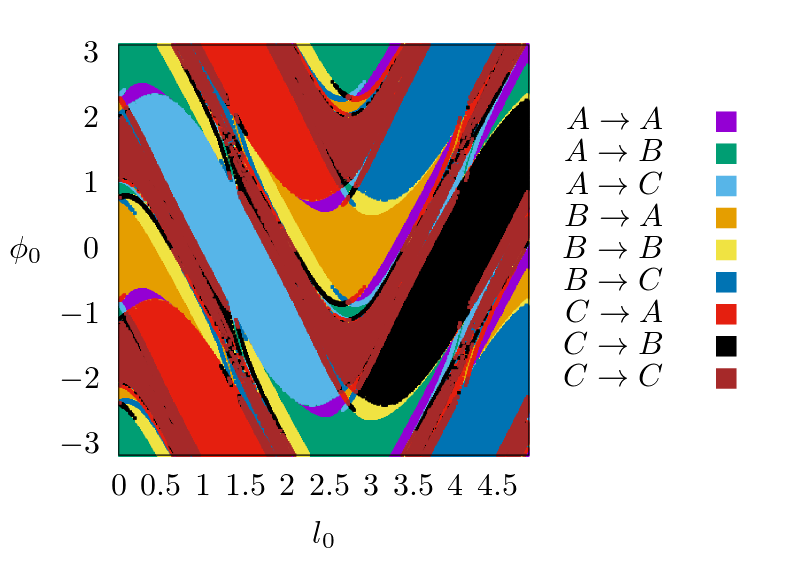}
(d)\includegraphics[scale=0.39]{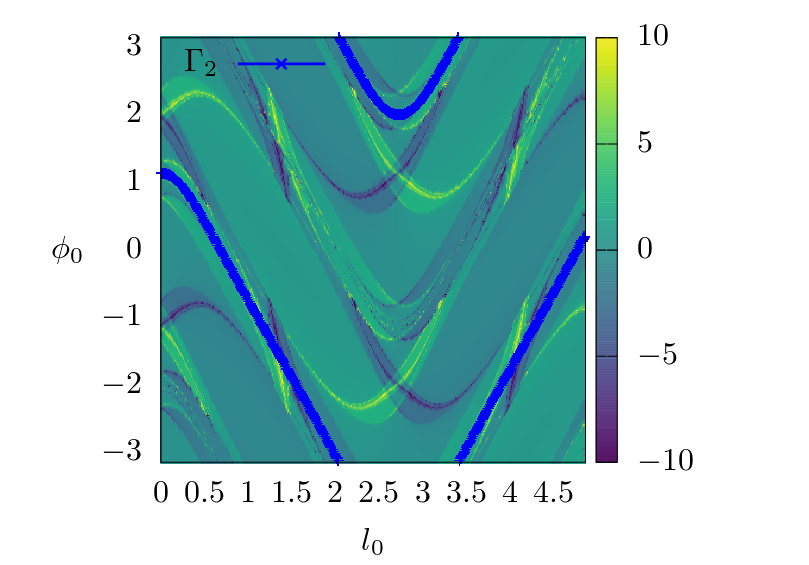}
(e)\includegraphics[scale=0.39]{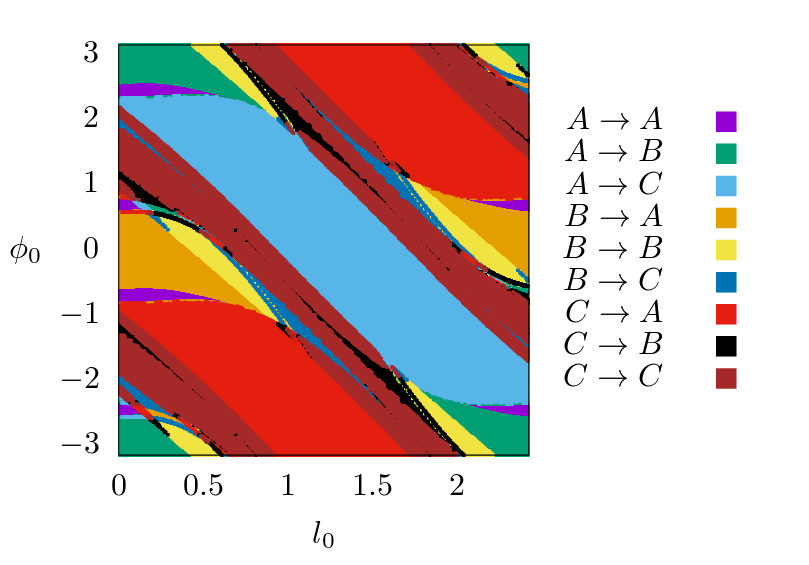}
(f)\includegraphics[scale=0.39]{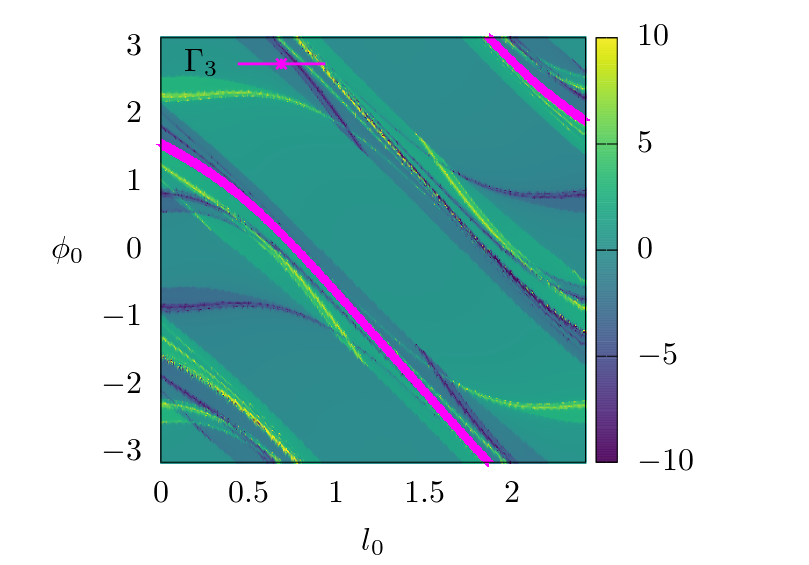}
\caption{Fate maps and $M^{\pm}_{S_0}$ evaluated on the dividing surface corresponding to the periodic orbits $\Gamma_1$, $\Gamma_2$, and $\Gamma_3$ for an energy before the bifurcation of $\Gamma_3$ ($E = 0.019$).}
\label{fig:ld_fm_ds_E_0019}
\end{figure}

\begin{figure}[htbp]
(a)\includegraphics[scale=0.39]{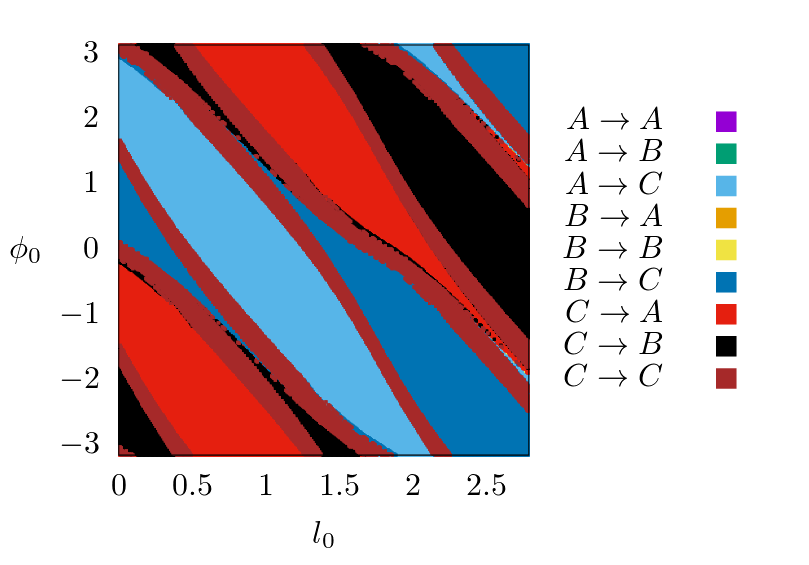}
(b)\includegraphics[scale=0.39]{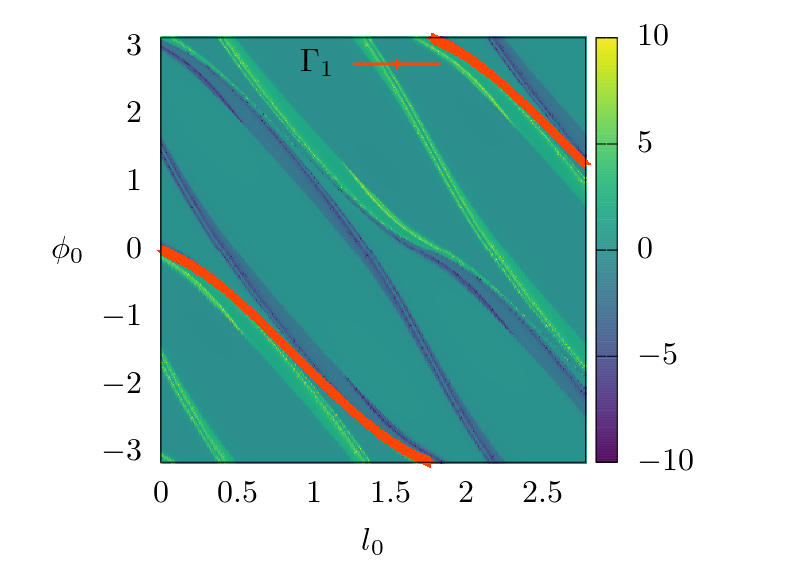}
(c)\includegraphics[scale=0.39]{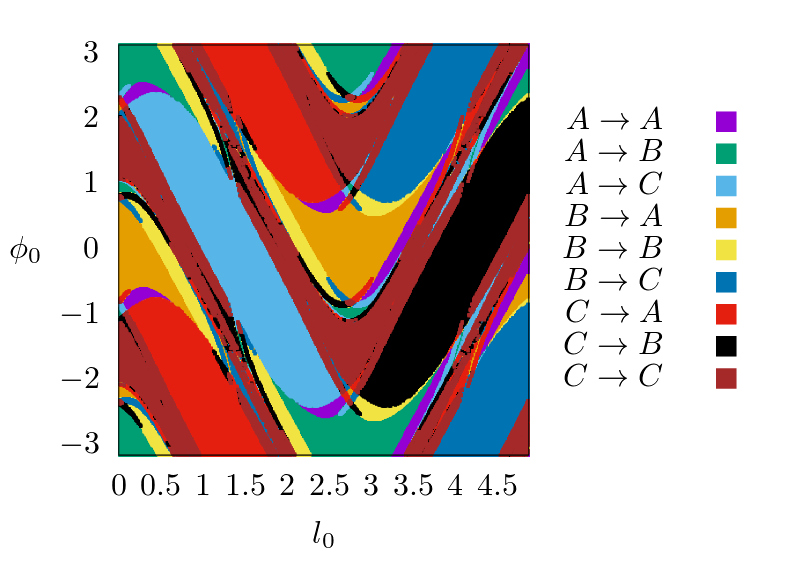}
(d)\includegraphics[scale=0.39]{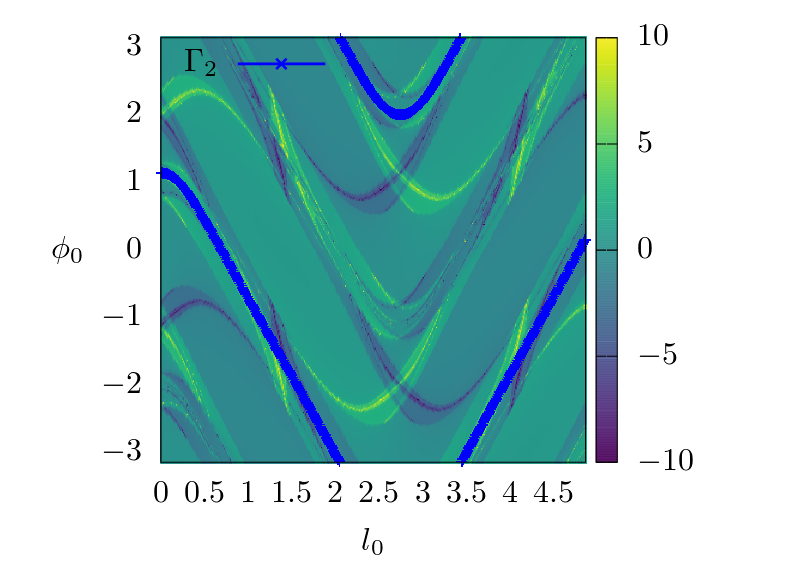}
\caption{Fate maps and $M^{\pm}_{S_0}$ evaluated on the dividing surface corresponding to the periodic orbits $\Gamma_1$ and $\Gamma_2$ for an energy after the bifurcation of $\Gamma_3$ ($E = 0.021$).}
\label{fig:ld_fm_ds_E_0021}
\end{figure}

\begin{figure}[htbp]
(a)\includegraphics[scale=0.39]{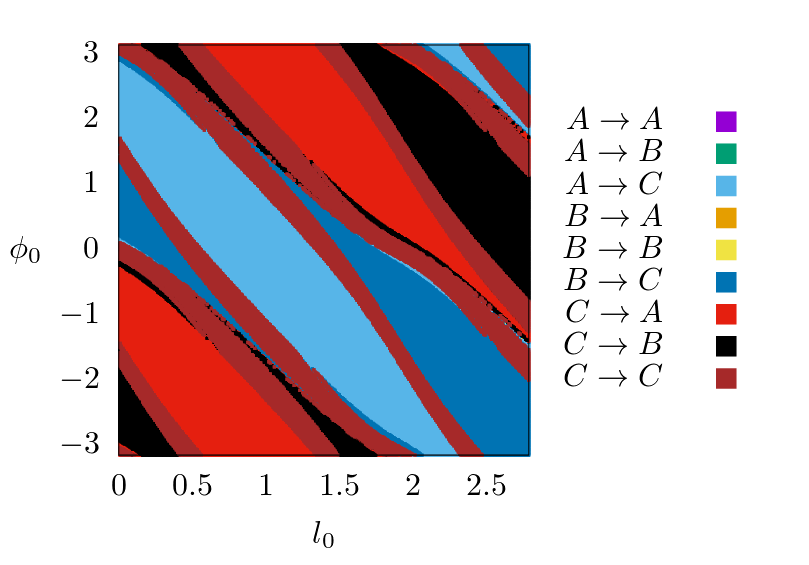}
(b)\includegraphics[scale=0.39]{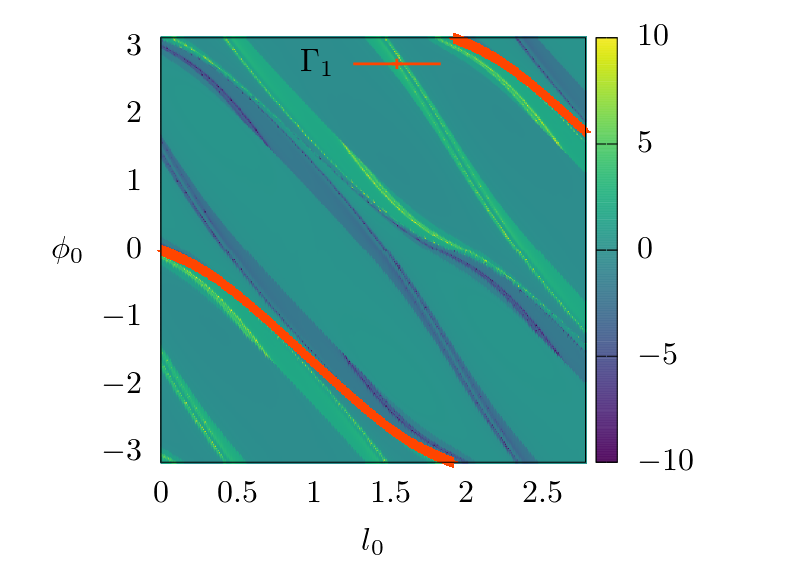}
(c)\includegraphics[scale=0.39]{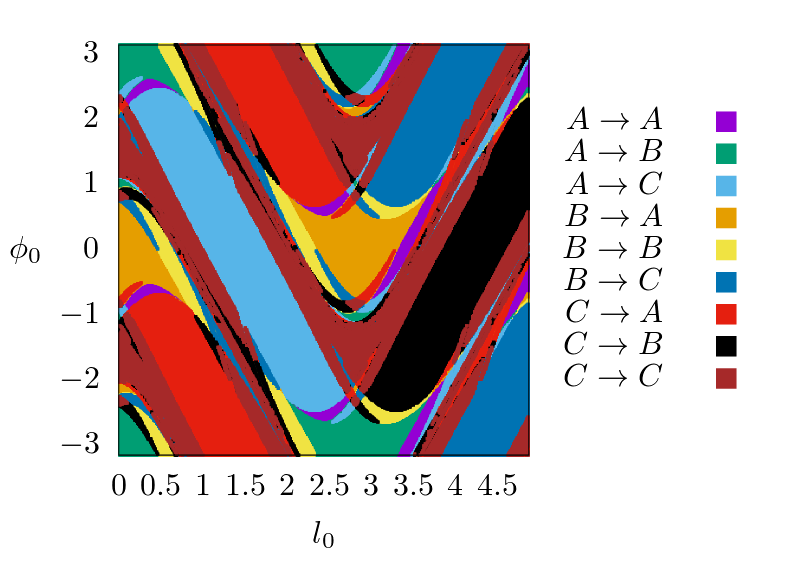}
(d)\includegraphics[scale=0.39]{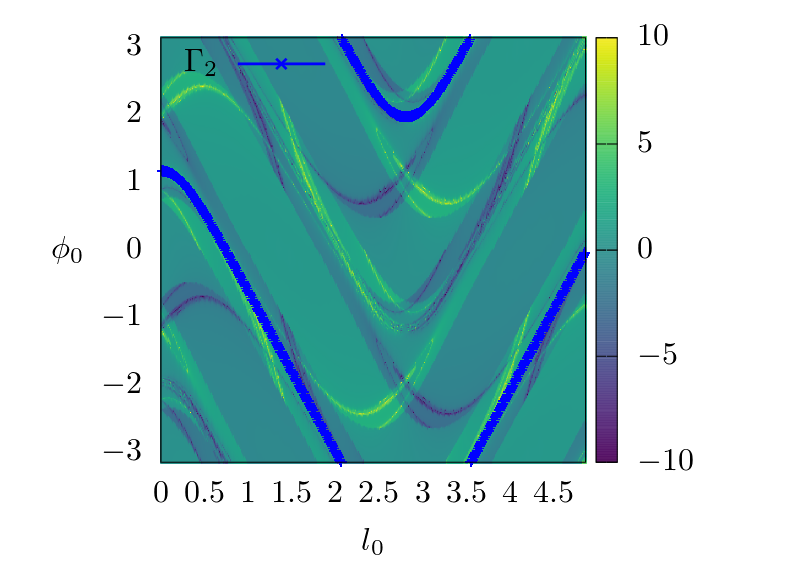}
\caption{Fate maps and $M^{\pm}_{S_0}$ evaluated on the dividing surface corresponding to the periodic orbits $\Gamma_1$ and $\Gamma_2$ for an energy before the bifurcation of $\Gamma_2$ ($E = 0.027$).}
\label{fig:ld_fm_ds_E_0027}
\end{figure}

\begin{figure}[htbp]
(a)\includegraphics[scale=0.39]{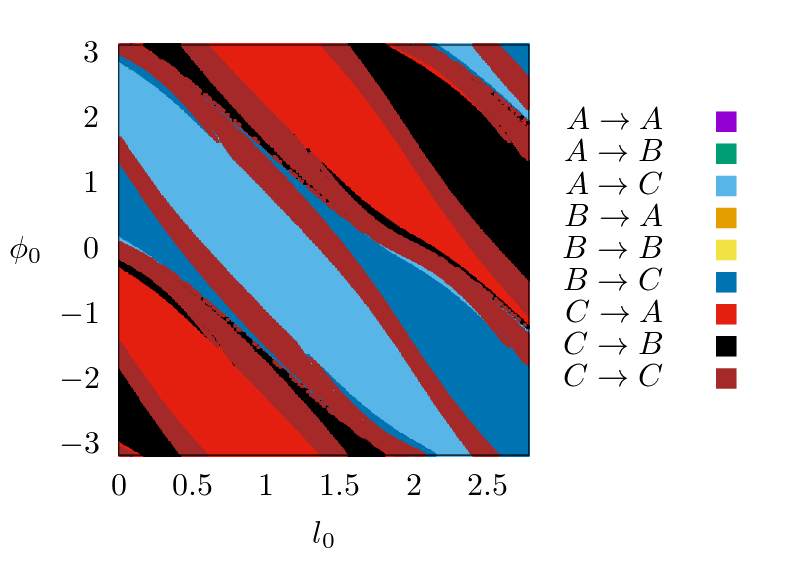}
(b)\includegraphics[scale=0.39]{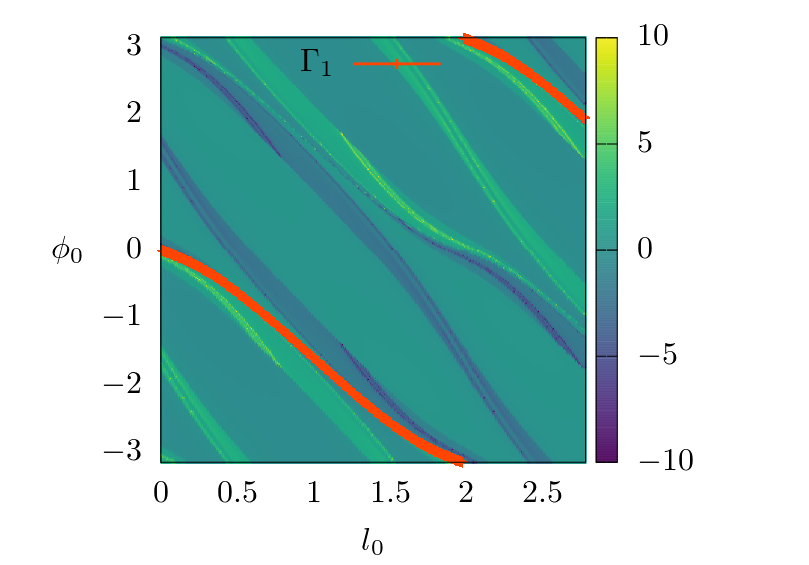}
\caption{Fate maps and $M^{\pm}_{S_0}$ evaluated on the dividing surface corresponding to the periodic orbits $\Gamma_1$ for an energy after the bifurcation of $\Gamma_2$  ($E = 0.03$).}
\label{fig:ld_fm_ds_E_003}
\end{figure}

\begin{figure}[htbp]
(a)\includegraphics[scale=0.39]{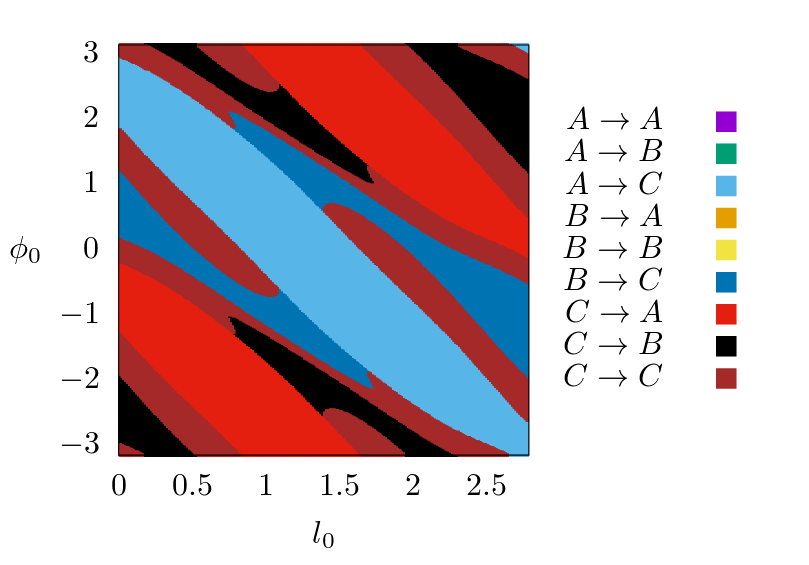}
(b)\includegraphics[scale=0.39]{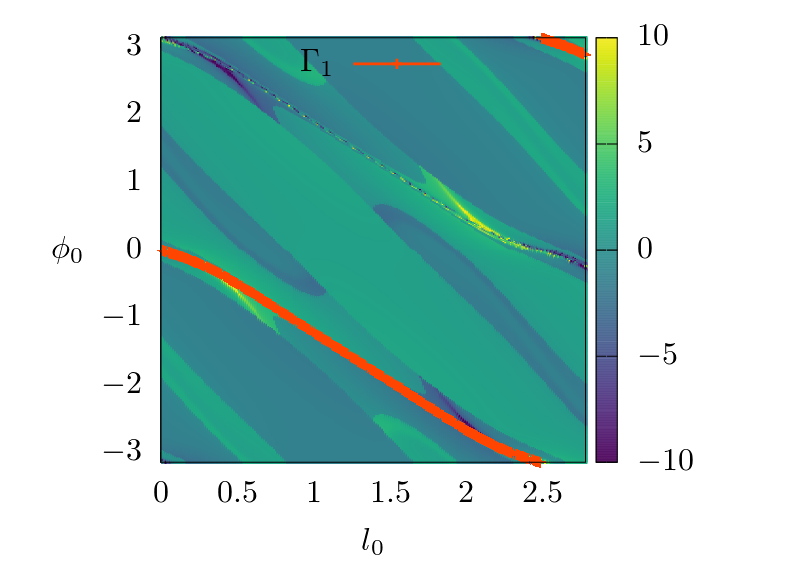}
\caption{Fate maps and $M^{\pm}_{S_0}$ evaluated on the dividing surface corresponding to the periodic orbit $\Gamma_1$ for an energy before the bifurcation of $\Gamma_1$ ($E = 0.065$).}
\label{fig:ld_fm_ds_E_0065}
\end{figure}

In order to distinguish between the stable and unstable manifolds on the next plots, we are going to consider a variant of $M_{S_0}$ like in \cite{GonzalezMontoya2020}. This variant is defined as $M^{\pm}_{S_0} =  M^{+}_{S_0} - M^{-}_{S_0}$. The abrupt changes in  $M^{+}_{S_0}$ correspond to the stable manifolds, and the abrupt changes in  $M^{-}_{S_0}$ correspond to the unstable manifolds. The Figures \ref{fig:ld_fm_ds_E_001}-\ref{fig:ld_fm_ds_E_0065} show the fate maps and the corresponding $M^{\pm}_{S_0}$ evaluated on the dividing surfaces of the periodic orbits $\Gamma_1$, $\Gamma_2$, and $\Gamma_3$ for different values of the energy. The dividing surface is parametrized by the angle $\phi_0 = \arctan{( p_{y0}/ p_{x0})}$ and the arclength $l_0$ defined by the value of the action $S_0$ along the periodic orbit in phase space. We consider values of $E$ before and after the bifurcation of the three families of hyperbolic periodic orbits $\Gamma_1$, $\Gamma_2$, and $\Gamma_3$. The fate maps show the transport between the three different regions $A$, $B$, and $C$ in configuration space, see Figs. \ref{fig:regions_A_B_C_potencial} and \ref{fig:periodic_orbits_roaming}. The trajectories start on the dividing surface and are calculated forwards and backwards until they reach any of the three regions. Clearly, comparing the fate maps and the $M^{\pm}_{S_0}$ plots show a good agreement between the boundaries of the regions with different fates and the abrupt jumps in the Lagrangian descriptor plots. These abrupt jumps have two origins: the stopping of the integration when the trajectories reach one of the regions $A$ or $B$, and the presence of stable and unstable manifolds of the hyperbolic periodic orbits. Hence, the stable and unstable manifolds combined with the boundaries of the regions $A$, $B$, and $C$ define all the boundaries of the regions with different fates on the constant energy manifold.

In Figures \ref{fig:ld_fm_ds_E_001}-\ref{fig:ld_fm_ds_E_0065} the size of the areas of the fate maps of the external periodic orbits do not change significantly after the bifurcation of the most internal periodic orbits, $\Gamma_3$. For example, the fate maps of the periodic orbits $\Gamma_1$ and $\Gamma_2$ after and before the bifurcation of the periodic orbit $\Gamma_3$ are similar, see Figs. \ref{fig:ld_fm_ds_E_0019} and \ref{fig:ld_fm_ds_E_0021}. The situation is similar for the other bifurcations, compare corresponding Figs. for the bifurcation of $\Gamma_2$ \ref{fig:ld_fm_ds_E_0027} and \ref{fig:ld_fm_ds_E_003}, and for the bifurcation of $\Gamma_1$ \ref{fig:ld_fm_ds_E_0065}.

The fate maps and Lagrangian descriptors for the lowest value of energy $E=0.01$, close to the threshold energy, are similar to the double Morse potential results, see Fig 11 in the reference \cite{GonzalezMontoya2020}. This similarity is related to the fact that the transport dynamics between the three regions is determined by the potential energy outside the regions $A$ and $B$. The potential energy surfaces outside regions $A$, $B$, and $C$ are similar for both systems and determine the dynamics and the fates of the trajectories. It is important to remark that the stable and unstable manifolds of a periodic orbit are asymptotic objects with codimension one, in the constant energy manifold, and their tangles form a partition of the entire phase space. Their presence in a specific phase space region is evident only when two trajectories nearby and on opposite sides of the stable manifold develop a different behaviour when they diverge close to the hyperbolic periodic orbit and following the opposite directions defined by the unstable manifold. However, trajectories may have a different fate due to the definition of the regions considered in the specific transport problem regardless of the tangles.
   
 \newpage

\newpage

\newpage

\newpage

 \newpage

\newpage

\section{Conclusions and Remarks}
\label{sec:conclusion}

In this work we have studied the phase space and transport in an unbounded potential energy surface constructed as the superposition of two van der Walls potential wells. This potential energy surface has similar properties to the double Morse potential energy surface analysed in previous work. These 2-dimensional models are an initial step to understand the key features that determine the transport and roaming in three and more dimensional models. In particular, we study the transport between the two potential wells and between one potential well and the asymptotic region for different values of the energy $E>0$. To visualise the phase space structure, we calculate a Lagrangian descriptor based on the action and fate maps.

The Lagrangian descriptors are useful tools for revealing the phase space objects that direct the dynamics and help us determine the values of the parameters for which bifurcation of hyperbolic periodic orbits occurs and the initial conditions to calculate those orbits. This trajectory-based approach is a convenient technique to find stable and unstable manifolds that do not require recurrent trajectories to any particular set, unlike Poincar\'e maps. This property facilitates the study of open systems where trajectories in the asymptotic region approach to the free particles trajectories and invariant manifolds of hyperbolic periodic orbits are unbounded.

In this example, the Lagrangian descriptor as a function of the energy allows us to find critical values for which the phase space structure has important changes. For instance, the bifurcation of stable and unstable invariant manifolds of hyperbolic periodic orbits or when the phase space has not transient chaos anymore, and the dynamics becomes simple. Due to its definition based on trajectories, Lagrangian descriptors are an advantageous simple tool for exploring the parameter space and finding where the hyperbolic periodic orbits disappear.

For the energies slightly bigger than the dissociation threshold, $E=0$, the phase space become unbounded, and some regions of the phase space are transient chaotic. There is an interval of the energy such that there are three important families of hyperbolic periodic orbits around the potential wells associated with the transport between potential wells. Their stable and unstable manifolds form boundaries of cylinders that direct the dynamics for trajectories to move from one region to another. The bifurcation energies for the three periodic orbits $\Gamma_3$, $\Gamma_2$, and $\Gamma_1$ are $E = 0.019$, $ 0.029$, and $0.659$ respectively, see the three corresponding points in Lagrangian descriptor plots on Fig. \ref{fig:ld_E_xy}. 
 
Studying the transition between bounded phase space and unbounded phase space is an interesting and challenging problem. In these systems, we can appreciate the large changes in the phase space structure and the creation of the most external periodic orbit. This periodic orbit is fundamental to understanding the unbounded system's dynamics and its differences with the bounded system. Their stable and unstable manifolds are the key element for understanding the transient chaos in the unbounded system.  
 
A remarkable difference between the double van der Waals potential energy surface and the double Morse one is the shape of the potential in the regions $A$ and $B$. The double Morse potential has a barrier in each region. Meanwhile, the double van der Waals has a minimum. In the double Morse, there is a hyperbolic periodic orbit oscillating on the $x$-axis between the two barriers contained in the regions $A$ and $B$ for $E>0$. In the double van der Waals system, the hyperbolic periodic orbit that oscillates on the $x$-axis exist only for $E<0$,  compare the central region of the Lagrangian descriptor plots in Fig. \ref{fig:ld_x_px} d) and Fig. 9 in \cite{GonzalezMontoya2020}. However, for energies close to the dissociation threshold $E=0$, the fate maps and Lagrangian descriptor plots evaluated on the dividing surfaces have a similar structure for both systems. This result is related to the shape of the potential energy surfaces outside the regions $A$, $B$. Both systems have a similar geometry outside those regions, a similar vector field that determines the dynamics, and similar stable and unstable manifolds define the boundaries of the fates maps. Notice that the trajectories only travel a finite time before reaching the boundaries that define the regions $A$, $B$, and $C$. Then some differences in the dynamics are not manifested for many trajectories that connect the three regions.

The central question in this work is the effect of the bifurcations of hyperbolic periodic orbits $\Gamma_1$, $\Gamma_2$, and $\Gamma_3$ on the transport between regions $A$, $B$, and $C$. Figs. \ref{fig:ld_fm_ds_E_001}-\ref{fig:ld_fm_ds_E_0065} show fate maps evaluated on their corresponding dividing surfaces for different values of $E$ close to bifurcation. Note that stable and unstable manifolds of the hyperbolic periodic orbits and the borders of the regions $A$ and $B$ bound regions with different transport fates in this problem. The number of trajectories arriving at each region is similar before and after each bifurcation. Also, the size of the larger tubes does not change significantly due to the bifurcation. In one way, the stable and unstable manifolds of the remaining periodic orbits retain the size of the larger tubes. The changes in the number of trajectories that arrive in each region due to the bifurcations are small. Another important fact independent of the boundaries between regions $A$, $B$, and $C$ is that for many trajectories travelling between these regions the vector field is very similar before and after these bifurcations. Therefore, some trajectories that connect those regions are not affected by the bifurcations.

\section*{Acknowledgements}

The authors would like to acknowledge the financial support provided by the EPSRC Grant No. EP/P021123/1 and the Office of Naval Research Grant No. N00014-01-1-0769.

\bibliography{cirque}
\end{document}